\documentclass{article}
\usepackage[utf8]{inputenc}
\usepackage{amsmath}
\usepackage{amsfonts}
\usepackage{optidef}
\usepackage{bbm}
\usepackage{parskip}
\usepackage{mathtools}
\usepackage{caption}
\usepackage{subcaption}
\usepackage{graphicx}
\usepackage[normalem]{ulem}
\usepackage{framed}
\usepackage{smartdiagram}
\usepackage{tikz}
\usetikzlibrary{arrows.meta,
                positioning}

\usepackage[
doi=false,isbn=false,url=false,eprint=false,
backend=biber,
style=numeric,
sorting=none,
giveninits=true,
maxbibnames=3
]{biblatex}
\renewbibmacro{in:}{}
\AtEveryBibitem{%
  \clearlist{language}%
  \clearfield{pages}
  \clearfield{note}
  \clearfield{month}
}
\title{Impact of limited temporal resolution on 4D Monte Carlo dose calculation for intensity modulated proton therapy}
\author{Ivar Bengtsson%
         \thanks{Optimization and Systems Theory, Department of
             Mathematics, KTH Royal Institute of Technology, SE-100 44
             Stockholm, Sweden (\texttt{ivarben@kth.se})}
        \thanks{RaySearch Laboratories AB, SE-104 30
             Stockholm, Sweden.}
  \and Erik Engwall\footnotemark[2]%
  \and Albin Fredriksson\footnotemark[2]%
  \and Lars Glimelius\footnotemark[2]%
  }
\date{\today}

\begin{document}

\maketitle

\begin{abstract}
The interplay between the beam delivery time structure and the patient motion makes 4D dose calculation (4DDC) important when treating moving tumors with intensity modulated proton therapy. 4DDC based on phase sorting of a 4DCT suffers from approximation errors in the assignment of spots to phases, since the temporal image resolution of the 4DCT is much lower than that of the delivery time structure.

In this study we investigate and address this limitation by a method which applies registration-based interpolation between phase images to increase the temporal resolution of the 4DCT. First, each phase image is deformed toward its neighbor using the deformation vector field that aligns them, scaled by the desired time step. Then Monte Carlo-based 4DDC is performed on both the original 4DCT (10 phases), and extended 4DCTs at increasingly fine temporal resolutions.

The method was evaluated on seven lung cancer patients treated with three robustly optimized beams, with simulated delivery time structures. Errors resulting from limited temporal resolution were measured by comparisons of doses computed using extended 4DCTs of various resolutions. The dose differences were quantified by gamma pass rates and volumes of the CTV that had dose differences above certain thresholds. The ground truth was taken as the dose computed using 100 phase images, and was justified by considering the diminishing effects of adding more images. The effect on dose-averaged linear energy transfer was also included in the analysis.

A resolution of 20 (30) phase images per breathing cycle was sufficient to bring mean CTV $\gamma$-pass rates for $3\%/3$mm ($2\%/2$mm) above 99\%. For the patients with well behaved image data, mean CTV $\gamma$-pass rates for $1\%/1$mm surpassed 99\% at a resolution of 50 images.

\end{abstract}

\section{Introduction}

Radiation therapy treatments of patients subject to intra-fractional motion require \textit{four-dimensional dose calculation} (4DDC) for accurate evaluation of the delivered dose distribution. 4DDCs are particularly important for particle treatments, as the constructive interference between the delivery time structure and the patient motion, known as the interplay effect, may distort the delivered dose compared to the dose that was planned \cite{lambert_intrafractional_2005, bert_quantification_2008, bert_motion_2011}.

A common method for performing 4DDCs is phase sorting \cite{bert_quantification_2008, grassberger_motion_2013, engwall_effectiveness_2018, engwall_4d_2018}, in which individual spots are assigned to their nearest phase in time based on their time of delivery, a 4DCT, and a patient motion model. Partial doses are then computed and accumulated on a common reference phase by use of deformable registration. Another method is the deforming-dose-grid algorithm \cite{boye_mapping_2013, zhang_online_2014}. Here, the motion and density changes affecting each dose calculation point in the reference dose grid are included directly in the analytical dose calculation. This is achieved by considering the changes in position and \textit{water equivalent thickness} (WET) for each calculation point, at the time of delivery of each individual spot. While phase sorting relies fully on the phases of an input 4DCT, the parameters in the deforming-dose-grid algorithm are obtained and managed separately.

A known limitation in 4DDC is the low temporal resolution of 4DCT images, which typically contain 8-10 volumetric images that each represent a different phase of the breathing cycle. Thus, for typical breathing periods in the order of a few seconds, the temporal resolution of 4DCT is a few hundred milliseconds. This resolution is notably coarse when juxtaposed with the delivery time of a single spot, typically around a few milliseconds. The effect of limited temporal resolution on the accuracy of 4DDC for \textit{intensity modulated proton therapy} (IMPT) has recently been subject to investigation. Seo et al. performed 4DDCs on a simulated water phantom moving perpendicular to the beam direction \cite{seo_temporal_2017}. They observed that a resolution of at least 14 (17) phase images was needed to achieve $\gamma$-pass rates $\geq 90\%$ ($1\%/1$mm) when phantom motion was 10 (20) mm, and a 4DDC using 1000 phases was taken as reference. It is reasonable to expect that finer temporal resolutions may be required for the non-homogeneous densities in a human patient. Later, Zhang et al. evaluated the impact of temporal resolution on the deforming-dose-grid algorithm for liver cases \cite{zhang_dosimetric_2019}. By interpolation of the dose grid motion at the known phases of the original 4DCT, they were able to simulate the patient motion at arbitrarily fine temporal resolution, while the WET for each calculation point at each delivery time instance was still derived from the corresponding WET nearest in time. When comparing doses computed with high and low temporal resolution 4DDC, the authors found that around $10 \%$ of the CTV voxels differed by $5 \%$ or more. However, the differences increased under certain treatment characteristics. In a follow-up study from the same research group, the investigation was extended to include lung data \cite{duetschler_limitations_2022}. There, the authors observed that the dosimetric impact of limited temporal resolution was lower than that of irregularities in the breathing motion. In particular, the authors highlighted the inaccuracies of basing the 4DDC on a single 4DCT acquired before treatment, combined with a motion model which assumes regular cycling of the available phases and disregards potential variations in breathing amplitude.

A first limitation of this analytical approach based on the deforming-dose-grid algorithm is the inability to address the limited time resolution of the density information contained in the WET. Another limitation is the accuracy of analytical dose calculation algorithms in heterogeneous media such as lung \cite{taylor_pencil_2017}. In the present paper, we therefore evaluate the effect of the temporal resolution on 4DDCs using phase sorting and Monte Carlo dose calculations. With deformable-registration-based image interpolation, we increase the temporal resolution of the 4DCT by generating additional phase images, thus including both motion and density information. We then perform Monte Carlo dose calculations on each phase image in the extended 4DCT, and accumulate the dose on a reference phase, just as in conventional phase sorting. In principle, our approach is similar to that of Rosu et al. \cite{rosu_how_2007}, who investigated a similar method but for photon beams. Their conclusion was that for photons, sufficient dosimetric accuracy was achieved with only a few phase images, or even the average image, from the breathing cycle. Finally, we also investigate the effect of limited temporal resolution on \textit{dose-averaged linear energy transfer} ($\text{LET}_\text{d}$) distributions in 4DDC. To the best of our knowledge, this is the first study of effects on $\text{LET}_\text{d}$ in this context.

\section{Method} \label{method}

\subsection{4DDC by phase sorting}

The 4DDC is based on the interplay evaluation tool available in RayStation (RaySearch Laboratories, Stockholm, Sweden). The tool applies conventional phase-sorting methodology which comprises three steps:

\begin{enumerate}
    \item Each spot is assigned to its nearest phase image based on the treatment delivery and the patient motion model.
    \item The partial dose resulting from the spots assigned to each phase image is computed using the Monte Carlo dose engine in RayStation.
    \item The partial doses are accumulated on the reference phase image by the ANACONDA DIR algorithm in RayStation \cite{weistrand_anaconda_2015}.
\end{enumerate}

\subsubsection{Delivery time structure and patient motion}

RayStation was used to generate treatment plans with a machine model of an IBA dedicated nozzle. A connection to the IBA ScanAlgo system was then used to simulate the delivery time structure. ScanAlgo takes ordered spot energies, positions and weights as input, and returns an ordered vector of spot start times representative of a clinical setting. As for patient motion, the employed model assumes repeated breathing cycles, in which the patient passes through each phase in the order indicated by the 4DCT, with equal duration in each phase. Thus, the patient motion can be fully determined by the start phase and the breathing period.

\subsubsection{Dose calculation}

For step two of the 4DDC, spots are assigned to phases based on their nearest phase image in time. The dose associated with spots on a particular phase $p$ in the set of phases $\mathcal{P}$ is computed using RayStation's Monte Carlo dose engine, with the maximal uncertainty set to $0.1 \%$ \cite{fracchiolla_clinical_2021}. The resulting partial dose is denoted $d_p$, with $d_p \in \mathbb{R}^m$ for a dose grid comprising $m$ voxels.

\subsubsection{Deformable registration}

The third step of the 4DDC requires that dose is accumulated on a reference phase of the 4DCT by mapping of the partial doses to the reference phase. The mapping is performed by deformably registering each phase image $p$ to the reference image, using the ANACONDA algorithm in RayStation. Although ANACONDA is a hybrid algorithm which can incorporate both image intensities and ROI contours, no controlling ROIs are used in the present paper due to the selection of lung cases for the numerical study, which typically exhibit sharp image contrast. The effect of deformations on delivered dose is represented by matrices $D_p$, such that $D_p \in \mathbb{R}^{m\times m}, \forall p \in \mathcal{P}$, with which the total dose $d(x)$, given spot weights $x$, is computed as:

\begin{equation}
    d(x) = \sum_{p \in \mathcal{P}} D_p d_p(x).
\end{equation}

\subsubsection{$\text{LET}_\text{d}$ accumulation}

The phase-sorting methodology allows also for 4D calculation of $\text{LET}_\text{d}$. For each phase $p$, a partial $\text{LET}_\text{d}$ distribution $l_p(x)$ can be additionally computed and deformed to the reference phase. Due to the dose averaging present in the static calculation of $\text{LET}_\text{d}$, the accumulation across phases is updated accordingly. For the accumulated $\text{LET}_\text{d}$ vector $l(x)$, each voxel's accumulated $\text{LET}_\text{d}$ is thus given by

\begin{equation}
    l^i(x) = \sum_{p \in \mathcal{P}} \Tilde{d}^i_p(x) \Tilde{l}^i_p(x) / \sum_{p \in \mathcal{P}} \Tilde{d}^i_p(x),
    \label{LETd_formula}
\end{equation}

where $\Tilde{d}^i_p(x), \Tilde{l}^i_p(x)$ are the \textit{deformed} dose and LET to voxel $i$, i.e. the $i$:th components of $D_p d_p(x)$ and $D_p l_p(x)$, respectively.

\subsection{Image interpolation} \label{interpolation_method}

Performing the 4DDC described above on a typical 4DCT data set suffers from the fact that the temporal resolution of the phase images in the 4DCT is lower than that of the spots. Thus, assigning spots to images on a nearest-neighbour basis in step 2 unavoidably results in approximation errors. To determine and to minimize such errors, we propose to create intermediate phase images by computing and interpolating the \textit{deformation vector field} (DVF) that aligns neighboring phase images, before deforming one neighbour toward the other using this interpolated DVF. The method is similar to what was done by Schreibmann et al. and Rosu et al. \cite{schreibmann_image_2006, rosu_how_2007}.

With a 4DCT as input, the method requires computing a deformable registration between neighbouring phase images, each resulting in a DVF defined by a set of deformation vectors $\{ v_i \}_{i=1}^{m},$ where $v_i \in \mathbb{R}^3, i = 1, \dots, m$. Such a vector field can be used to generate the deformation of the target image that best matches the reference image, by assigning to point $y_i$ in the reference the intensity at point $y_i + v_i$ in the target, determined by trilinear interpolation. The interpolation method for generating an image representing the patient state after an arbitrary fraction $\alpha$ of the transition between neighbouring phase images, denoted $R$ and $T$, with $\alpha \in (0, 1)$, can be summarized as:

\begin{enumerate}
    \item [Input :]  Reference image $R$, target image $T$, and scaling parameter $\alpha$.
    \item Compute the DVF from $R$ to $T$.
    \item Rescale the DVF by $\alpha$ and deform $T$ by the scaled DVF, generating the interpolated image $I_\alpha$.
    \item Rescale the original DVF by $1 - \alpha$ and use it to propagate the contours from $R$ to $I_\alpha$.
    \item [Output:] Interpolated image $I_\alpha$ with corresponding contours.
\end{enumerate}

\subsection{Increased temporal resolution in 4DCT}

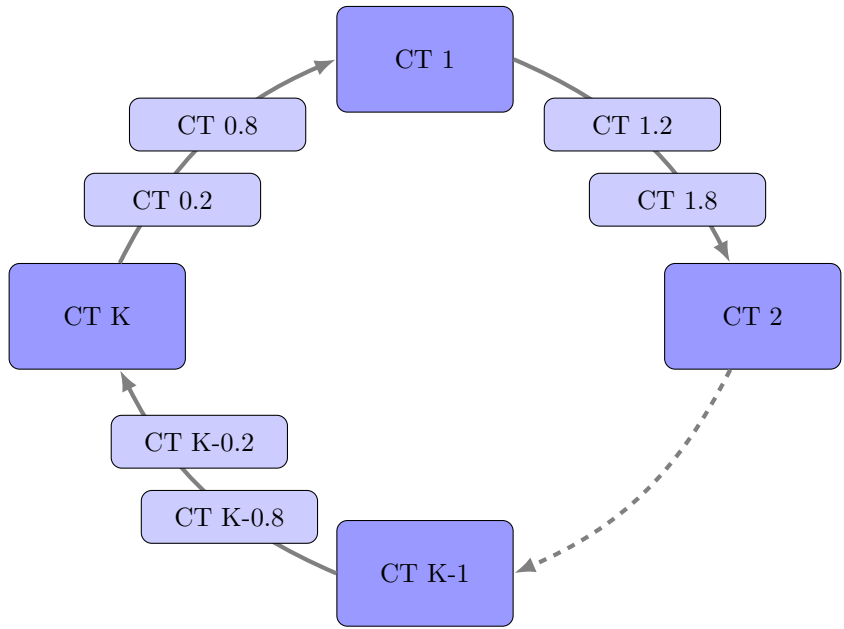
\begin{figure}
    \centering

\begin{tikzpicture}[
node distance = 2.0cm and 2.0cm,
block/.style = {rectangle, draw, fill=blue!40,
                text width=6em, text centered, rounded corners, minimum height=4em},
smallblock/.style = {rectangle, draw, fill=blue!20,
                text width=6em, text centered, rounded corners, minimum height=2em},
every edge/.style = {draw=gray, -{Latex[scale=0.8]}, ultra thick, bend angle=20},
                    ]
    \node [block] (1) {CT 1};
    \node [block, below right= of 1] (2) {CT 2};
    \node [block, below left= of 2] (3) {CT K-1};
    \node [block, above left= of 3] (4) {CT K};
    
    \path   (1.east) edge [bend left] (2)
            (2) edge [dashed, bend left] (3.east)
            (3.west) edge [bend left] (4)
            (4) edge [bend left] (1.west);

    \node [smallblock, below right=-0.2cm and 0.4cm of 1] (5) {CT 1.2};
    \node [smallblock, below right=0.8cm and 1.0cm of 1] (6) {CT 1.8};

    \node [smallblock, below left=-0.2cm and 0.4cm of 1] (5) {CT 0.8};
    \node [smallblock, below left=0.8cm and 1.0cm of 1] (6) {CT 0.2};

    \node [smallblock, below right=0.6cm and -1.0cm of 4] (5) {CT K-0.2};
    \node [smallblock, below right=1.6cm and -0.6cm of 4] (6) {CT K-0.8};
    
    \end{tikzpicture}
    
    \caption{A schematic representation of the model for increasing the temporal resolution in a 4DCT, exemplified with $\alpha = 0.2, 0.8$.}
    \label{extended4dct}
\end{figure}

The interpolation method described above allows for modelling of the patient state at time points between known phases. From a given 4DCT consisting of $K$ phases, an extended 4DCT consisting of the original phases and a multitude of interpolated phases is created in the following way: For each pair of consecutive phases in the 4DCT, a series of images $I_\alpha$, corresponding to $n$ evenly spaced values of $\alpha$, are created by interpolation. For this purpose, the DVF is computed from $\text{CT}_k$ to $\text{CT}_{k+1}$, for $k = 1, \dots, K-1$, and from $\text{CT}_K$ to $\text{CT}_1$. Each desired new phase is then created by the method in \ref{interpolation_method} and added to the 4DCT; the result is an extended 4DCT. A schematic illustration of the method is shown in Figure \ref{extended4dct}.

\subsection{Numerical cases}

\subsubsection{Patient data}

Seven patients from a data set uploaded to The Cancer Imaging Archive (TCIA) \cite{clark_cancer_2013} were used in the numerical experiments. The full data set consists of 20 non-small cell lung cancer patients and was described in detail by Hugo et al. \cite{hugo_longitudinal_2017}. The seven patients included in the present study were selected on the same basis as in the study by Engwall et al. \cite{engwall_effectiveness_2018} : to guarantee the existence of ROI contours, breathing period information and sufficiently challenging patient motion. Patient motion and breathing characteristics are presented in Table \ref{patients}.

\begin{table}[]
    \centering
    \begin{tabular}{c|c|c}
    \hline
        ID & Motion (cm) & Mean breathing period (s) \\ \hline
        P101 & 0.74 & 3.7 \\ \hline
        P103 & 0.47 & 3.3 \\ \hline
        P104 & 0.6 & 3.5 \\ \hline
        P110 & 0.53 & 7.0 \\ \hline
        P111 & 1.22 & 3.2 \\ \hline
        P114 & 0.97 & 3.2 \\ \hline
        P115 & 0.37 & 6.6 \\ \hline
    \end{tabular}
    \caption{Magnitude of motion as well as the mean breathing period for each patient. Motion is measured as the mean displacement vector length within the ITV, when registering the maximum expiration phase to the maximum inspiration phase. The mean breathing period was computed from the breathing periods during image acquisition.}
    \label{patients}
\end{table}

\subsubsection{Numerical experiments}

Treatment planning for each patient was performed by 4D-robust optimization in RayStation, with robustness settings of $5$ mm uncertainty in setup and $3 \%$ in range. All 10 phase images of the original 4DCT of each patient were included, resulting in a total of 210 scenarios used in optimization. This robust optimization minimizes the objective in the worst case over all the scenarios, without considering motion effects.

4DDCs were then run with with two primary intentions. The first intention was to determine the error of the 4DDC as a function of temporal resolution, as well as the resolution needed for sufficient accuracy in the 4DDC. To do so, a comparison was made between doses computed on extended 4DCTs comprising an increasing number of phase images. The lowest resolutions considered were those corresponding to each patient's original 10 4DCT phases, reflecting the conventional resolution in phase-sorting 4DDC. Dose calculations were then performed for extended 4DCTs of increased temporal resolution, in increments of 10 interpolated phase images evenly distributed across the breathing cycle. The highest resolutions corresponded to 100 phase images and were deemed sufficiently accurate based on the observations. For all comparisons, dose differences were measured with respect to a given reference dose, which depended on the purpose of the comparison. The metrics considered for comparing any two dose distributions were then $\gamma$-pass rates in the CTV as well as $V_{\text{dosediff} > \beta \%}$, the relative volume  of the CTV surpassing a voxel-wise dose difference threshold $\beta \%$, relative to the prescribed dose. Both metrics were considered useful since they were each used in the studies by Seo et al. \cite{seo_temporal_2017} and Zhang et al. \cite{zhang_dosimetric_2019}. Here, the start phase of the patient motion was held constant across all 4DDCs, and the breathing period was taken from Table \ref{patients}. The second intention was to juxtapose the error induced by limitations in temporal resolution with that arising from other error sources, in particular that induced by an offset of the start phase or a change in breathing period. To increase the number of data points available for comparison, each error source was investigated by comparison of 4DDCs with the start phase varying among the phases in the 4DCT.

In addition to the physical doses, $\text{LET}_\text{d}$ distributions were also computed for all 4DDCs. The metric used for comparing any two $\text{LET}_\text{d}$ distributions was $V_{\text{LET$_\text{d}$-diff} > \beta}$, defined as the relative volume of the high-dose region (receiving $>10\%$ of the prescription) with voxel-wise $\text{LET}_\text{d}$ differences greater than a threshold $\beta$.

\section{Results} \label{results}
 
\subsection{Convergence to a limit dose at increased temporal resolution} \label{convergence}

\begin{figure}

\subfloat[Dose differences in the CTV measured by $V_{\text{dosediff} > \beta \%}$, the relative volume  of the CTV surpassing a voxel-wise dose difference threshold $\beta \%$, relative to the prescribed dose, with $\beta \in \{ 1, 2, 3 \}$. \label{V dose diffs}]{%
\includegraphics[clip,width=\textwidth]{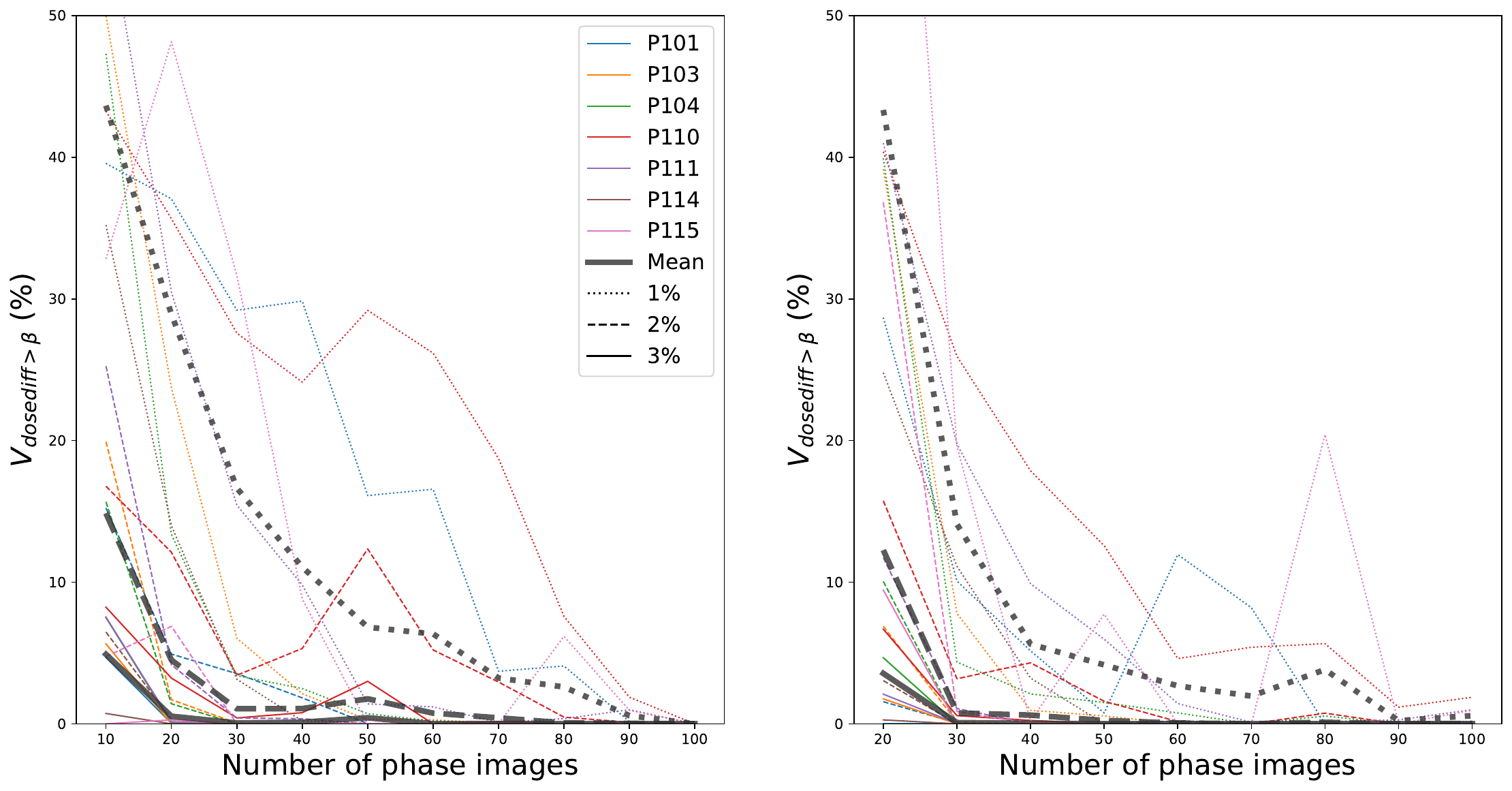}%
}

\subfloat[Dose differences in the CTV measured by $\gamma$-pass rates in the CTV for $1\%/1mm$, $2\%/2mm$, and $3\%/3mm$.
\label{gamma dose diffs}]{%
\includegraphics[clip,width=\textwidth]{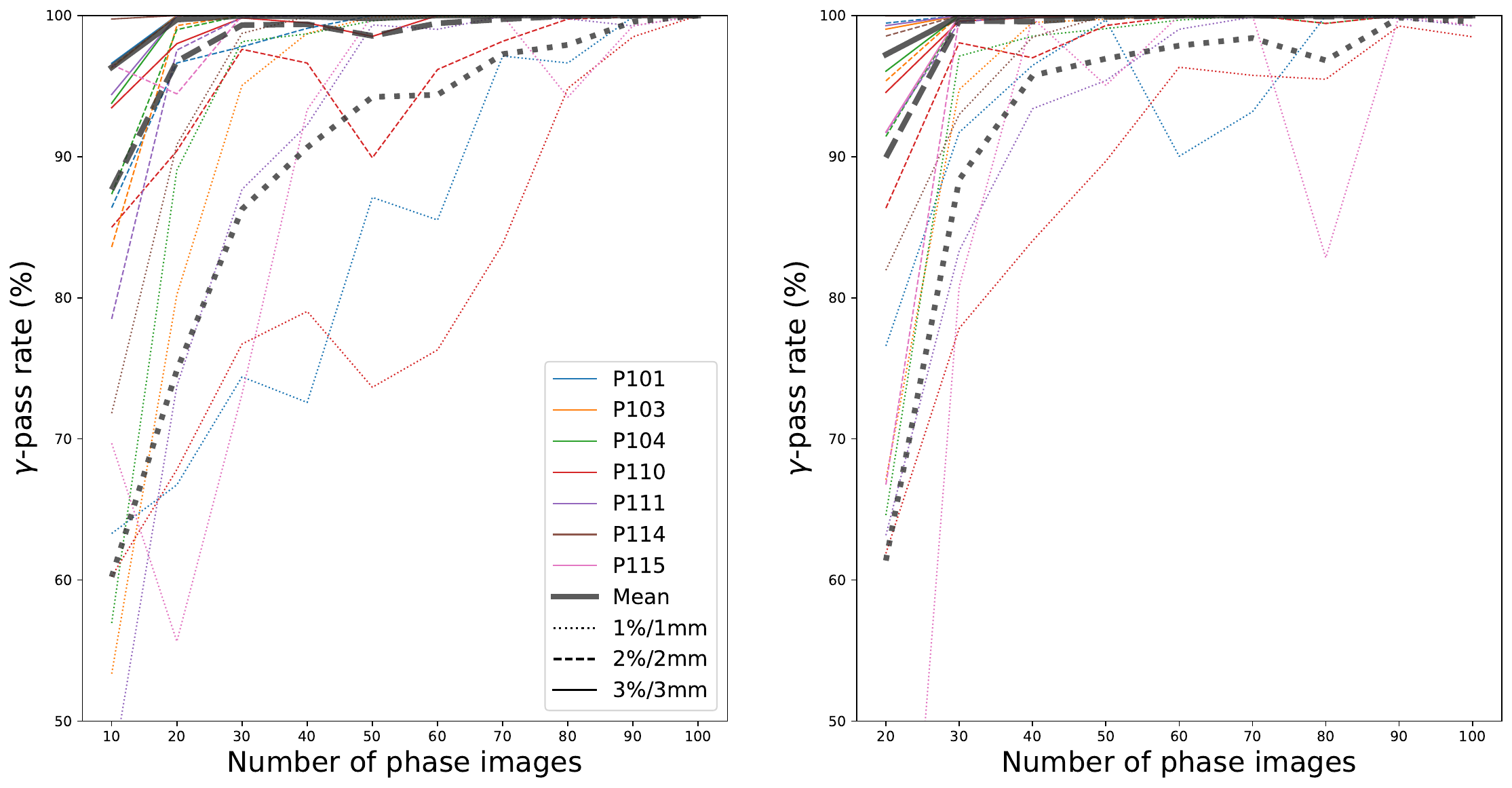}%
}
\caption{CTV dose differences as 4DCT temporal resolution is increased. On the left, differences from $d^i$ are measured with respect to the dose $d^{100}$, computed at maximal temporal resolution. On the right, differences from $d^i$ are measured with respect to the reference $d^{i-10}$, the dose computed at the previous temporal resolution. The patient label is indicated by the line color and the metric thresholds by the line style. The means across the patient group are drawn in thick gray.
}
\label{convergence_figures}
\end{figure}

Figure \ref{convergence_figures} shows dose differences between the dose $d^i$ computed with the temporal resolution of a 4DCT comprising $i$ phase images, and a reference dose $d^j$. In the panels on the left the reference is the dose $d^{100}$, computed on the extended 4DCT of highest resolution (100 phases), which is considered as the ground truth. As motivation for this choice, differences between consecutive doses at each increment in temporal resolution are shown in the panels on the right. There, the reference is the dose $d^{i-10}$, computed on the the extended 4DCT comprising 10 fewer phase images than that of $d^i$. This way, the effect of each additional increase of the temporal resolution on the 4DDC was measured. The dose differences tending to zero on the left in Figures \ref{V dose diffs} and \ref{gamma dose diffs} imply that the doses at increasing temporal resolutions of the 4DCT are increasingly similar to $d^{100}$. The mean dose differences across the seven patients decrease fast for the less strict metrics. $\gamma$ -pass rate is above 99$\%$ already at 20 phase images for $3\%/3\text{mm}$, and at 30 phase images for $2\%/2\text{mm}$. Likewise, $V_{\text{dosediff} > 3 \%}$ is below 1$\%$ at 20 phase images, and $V_{\text{dosediff} > 2 \%}$ is approximately 1$\%$ at 30 phase images. Measured by the the stricter metrics ($\gamma \text{ with } 1\%/1\text{mm}$, $V_{\text{dosediff} > 1 \%}$) however, $d^i$ tends to $d^{100}$ less fast. Most notably, patients P101, P110 and P115 do not exhibit as clear convergence as the patients P103, P104, P111 and P114. This discrepancy is addressed in the discussion as well as in Appendix \ref{appendixA}. However, when these patients are disregarded, 50 phase images are sufficient to get the mean $\gamma \text{-pass rate with } 1\%/1\text{mm}$ above 99$\%$, as well as the mean $V_{\text{dosediff} > 1 \%}$ below 1$\%$.

\begin{figure}
\centering
\subfloat{%
\includegraphics[clip,width=0.45\textwidth]{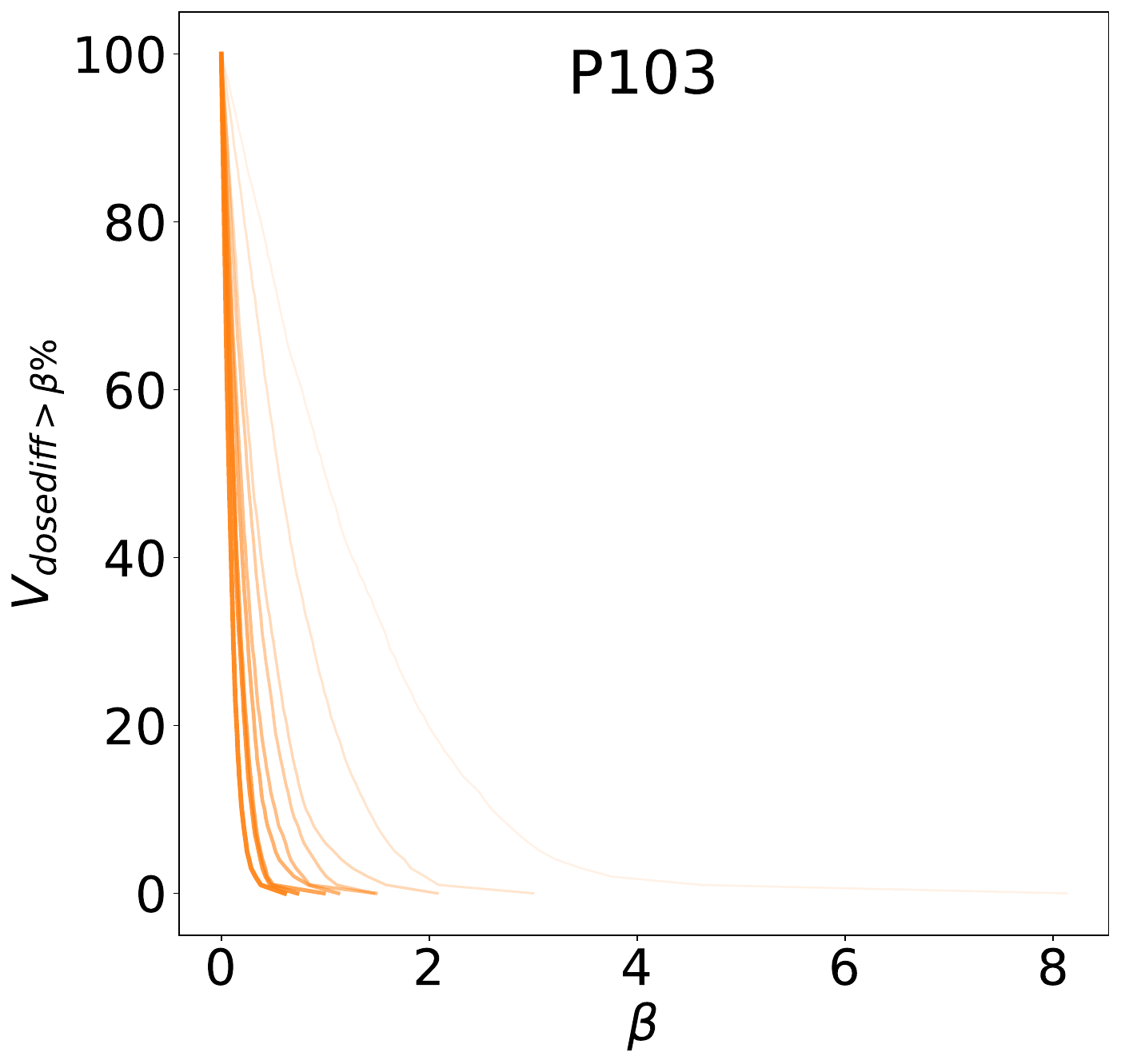}%
}
\subfloat{%
\includegraphics[clip,width=0.45\textwidth]{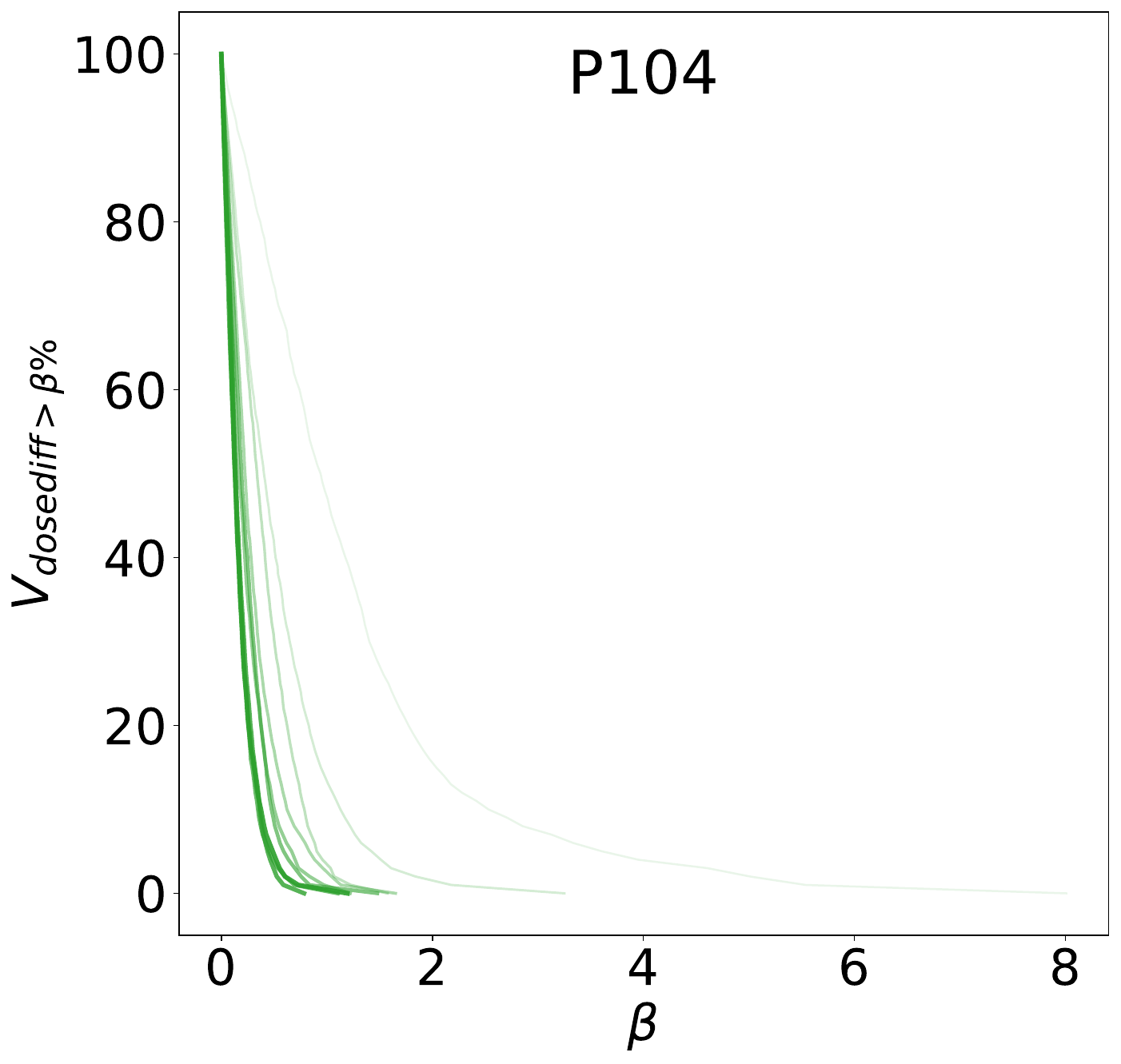}%
}

\subfloat{%
\includegraphics[clip,width=0.45\textwidth]{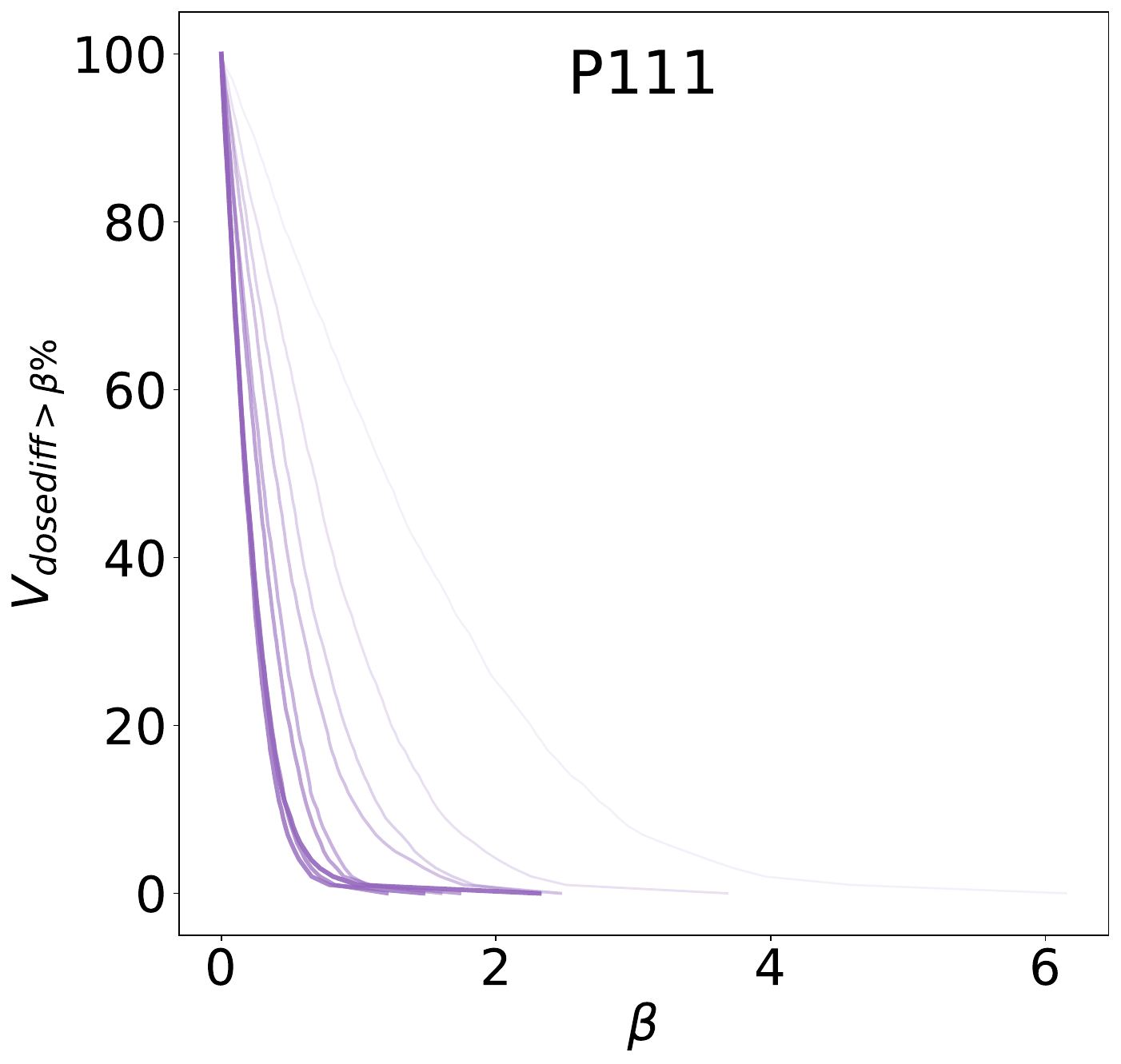}%
}
\subfloat{%
\includegraphics[clip,width=0.45\textwidth]{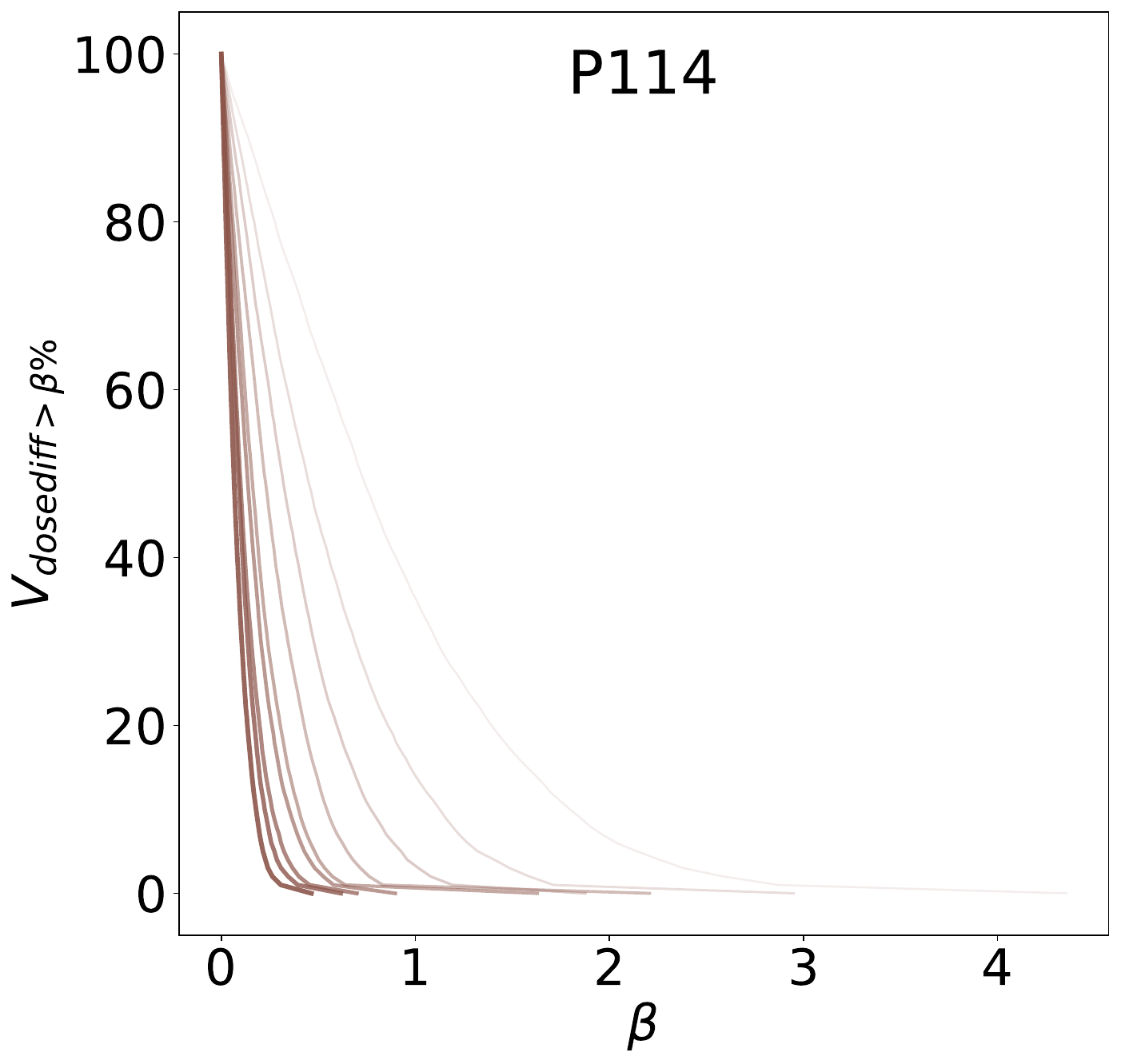}%
}

\subfloat{%
\includegraphics[clip,width=0.3\textwidth]{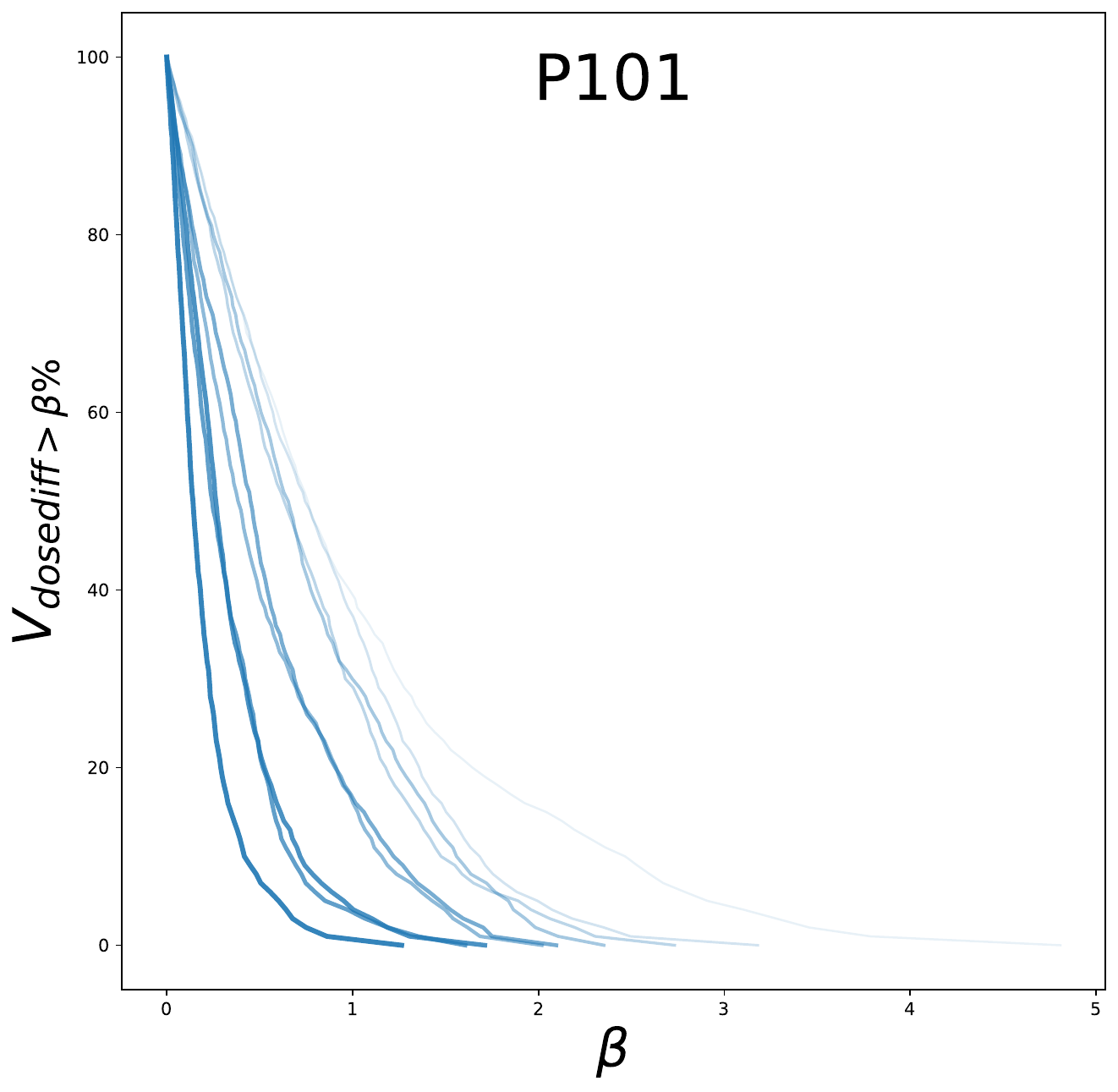}%
}
\subfloat{%
\includegraphics[clip,width=0.3\textwidth]{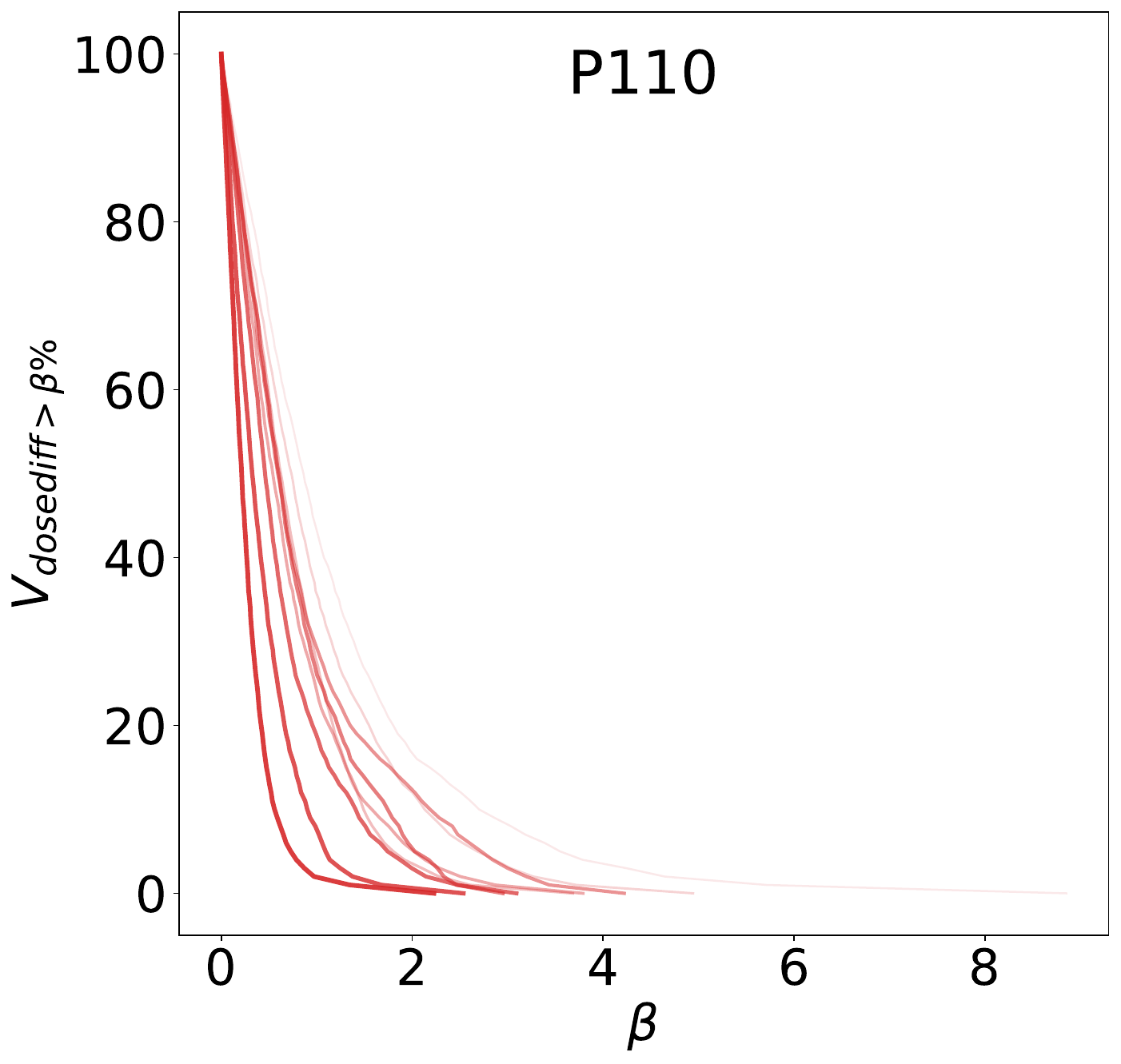}%
}
\subfloat{%
\includegraphics[clip,width=0.3\textwidth]{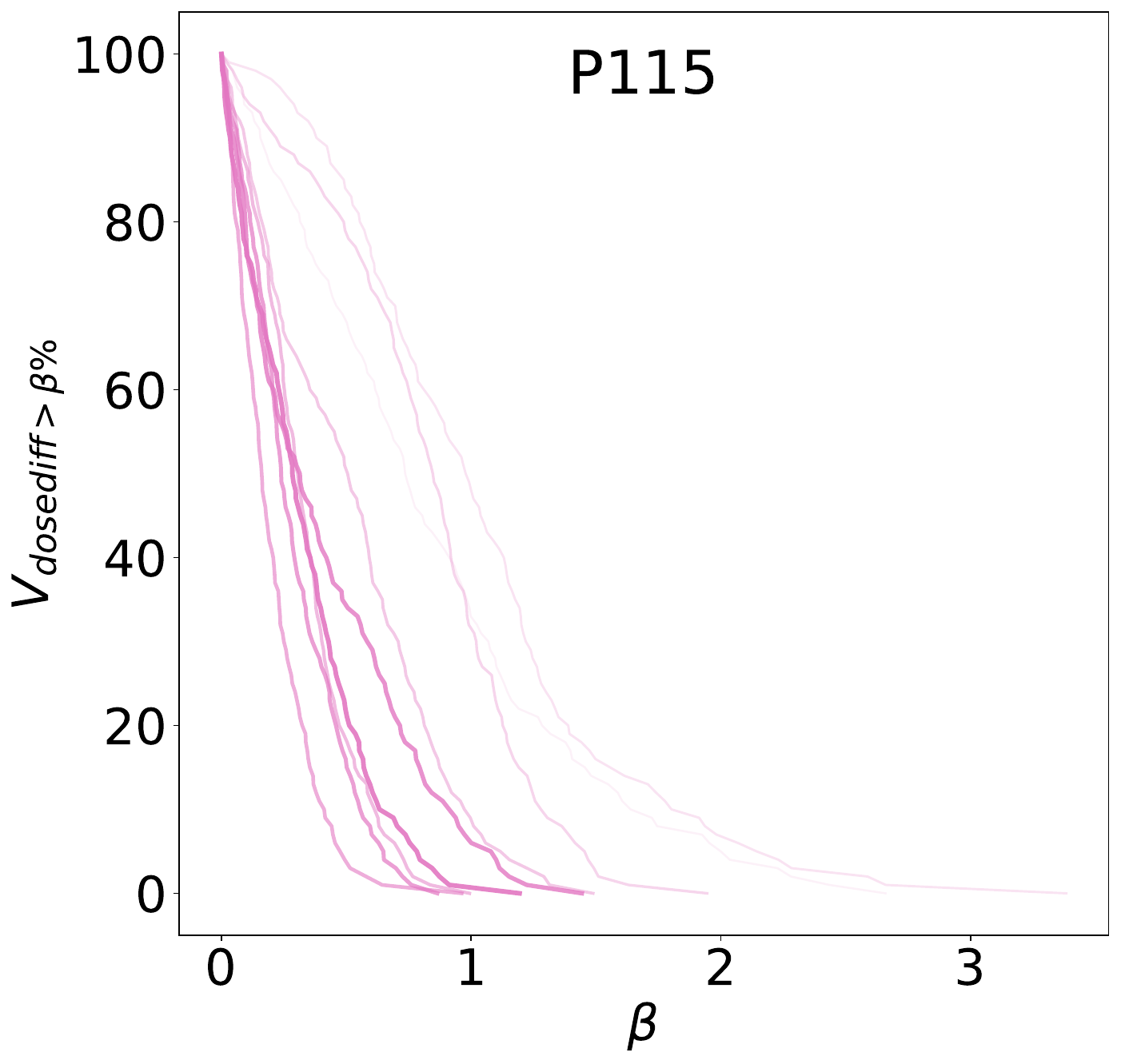}%
}

\caption{Dose differences histograms for each patient and temporal resolution. Each curve represents the dose difference when $d^i$, computed at the resolution with $i$ images, is compared to $d^{100}$. A point on any curve has coordinates $(\beta, V_{\text{dosediff} > \beta \%})$. Greater opacity and line width indicate finer resolution.}
\label{histograms}

\end{figure}

For additional detail, and removing the depency of the results on the choice of the threshold $\beta$, Figure \ref{histograms} displays the cumulative histograms of the dose difference distributions at each resolution, when $d^{100}$ is used as reference. More precisely, each point in a histogram has coordinates $(\beta, V_{\text{dosediff} > \beta \%})$. Like in Figure \ref{convergence_figures}, the convergence results are apparent for patients P103, P104, P111, and P114, whilst more ambiguous for P101, P110, and P115.

\begin{figure}
    \centering
    \includegraphics[width=\textwidth]{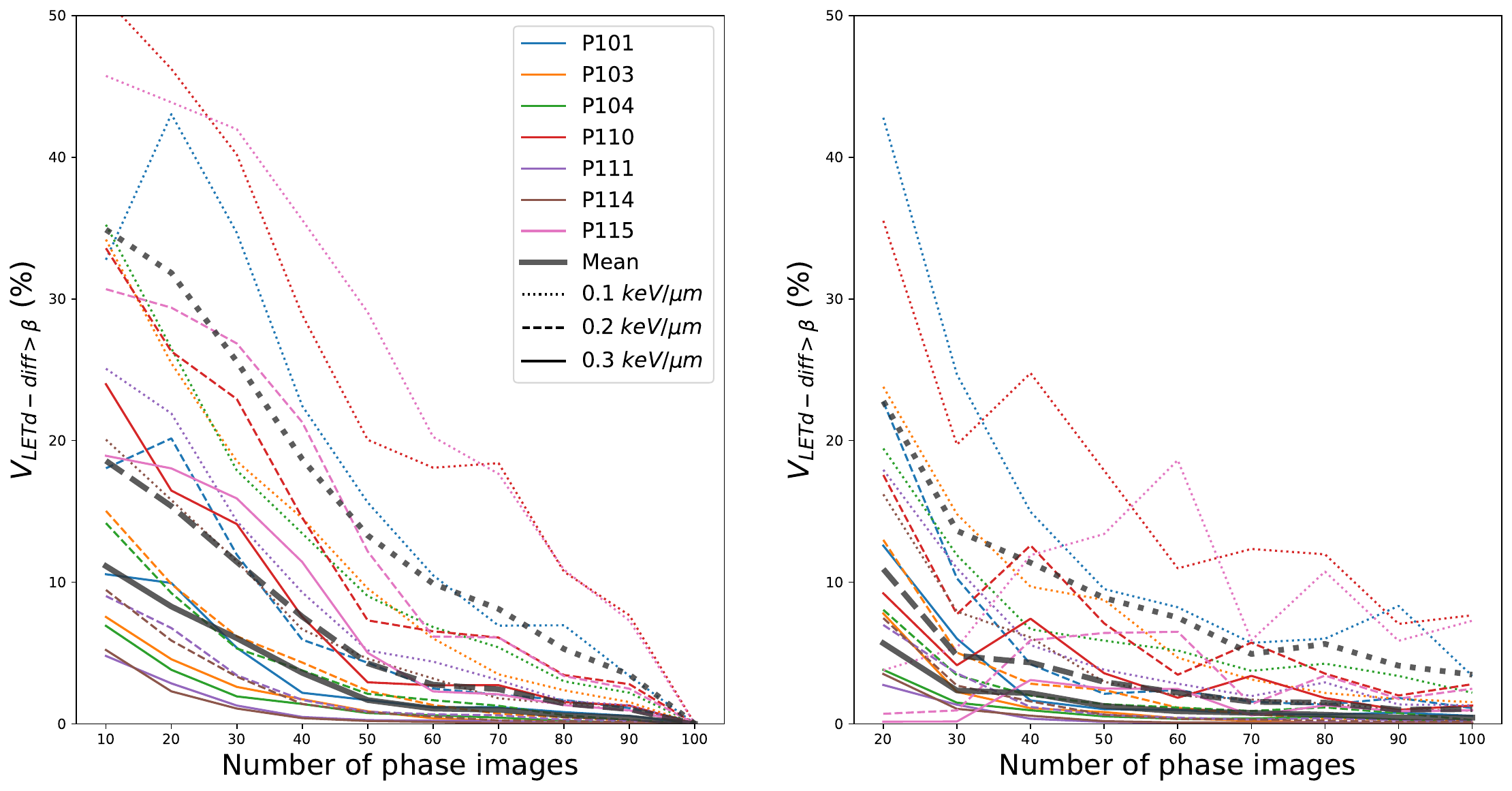}
    \caption{$\text{LET}_\text{d}$ differences as 4DCT temporal resolution is increased, measured by $V_{\text{LET$_\text{d}$-diff} > \beta }$ with $\beta$ = 0.1, 0.2, 0.3 keV/$\mu \text{m}$. The volume is measured relative to the total volume of the high-dose region ($>10\%$ of prescription). On the left and right, differences from $l^i$ are measured with respect to $l^{100}$ and $l^{i - 10}$, respectively.}
    \label{LETd}
\end{figure}

Figure \ref{LETd} shows results analogous to those in Figure \ref{V dose diffs} but with respect to $\text{LET}_\text{d}$. The value on the vertical axis of each data point $l^i$ is the difference between $l^i$ when compared to either $l^{100}$ (left) or $l^{i-10}$ (right). Convergence to a limit $\text{LET}_\text{d}$-distribution is not as apparent as for physical dose, as suggested by the panel on the right of Figure \ref{LETd}, although the mean difference has a notable negative trend.

\begin{figure}[]

\subfloat[$d^{10} - d^{100}$]{%
\includegraphics[trim={3cm 0 4cm 0},clip,width=0.33\textwidth]{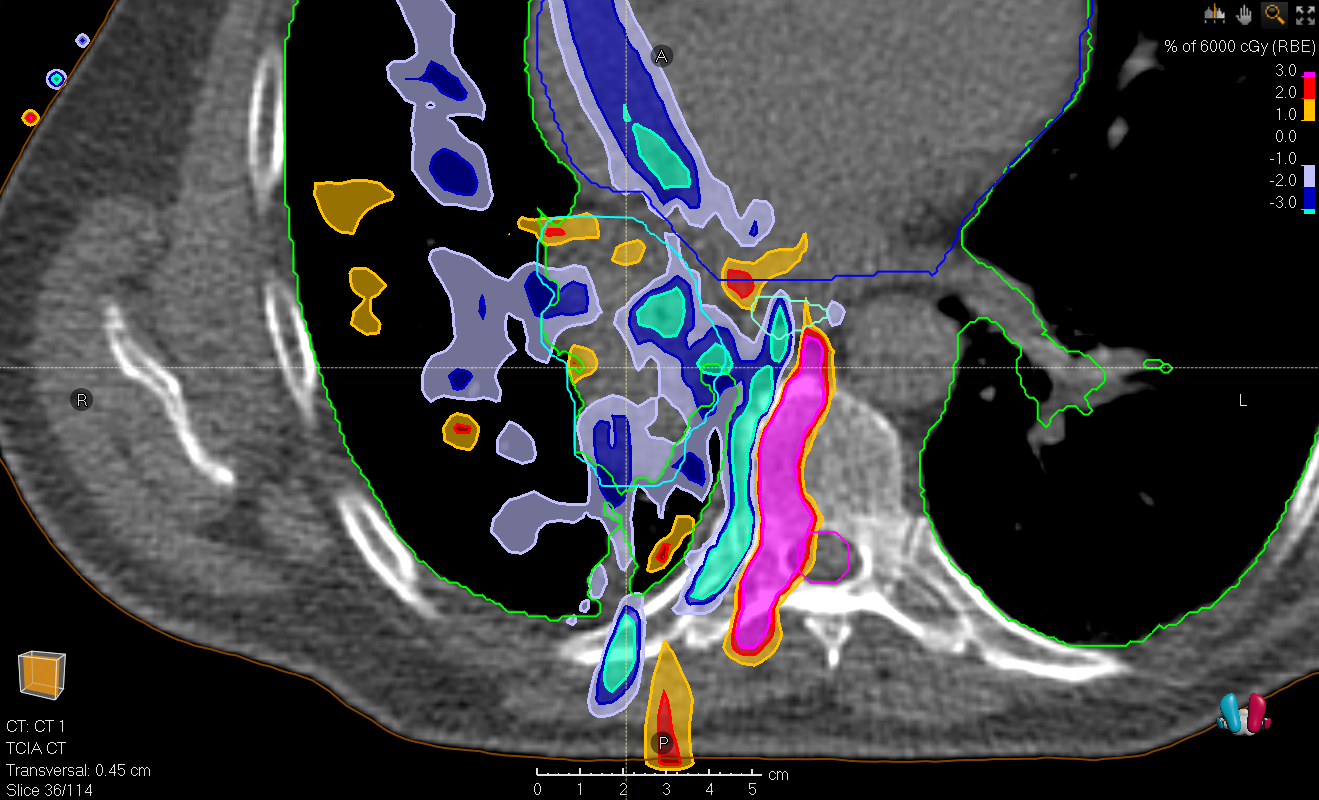}%
\llap{{%  move next graphics to top right corner
\includegraphics[height=2.5cm]{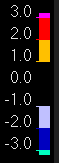}%
}}
}
\subfloat[$d^{20} - d^{100}$]{%
\includegraphics[trim={3cm 0 4cm 0},clip,width=0.33\textwidth]{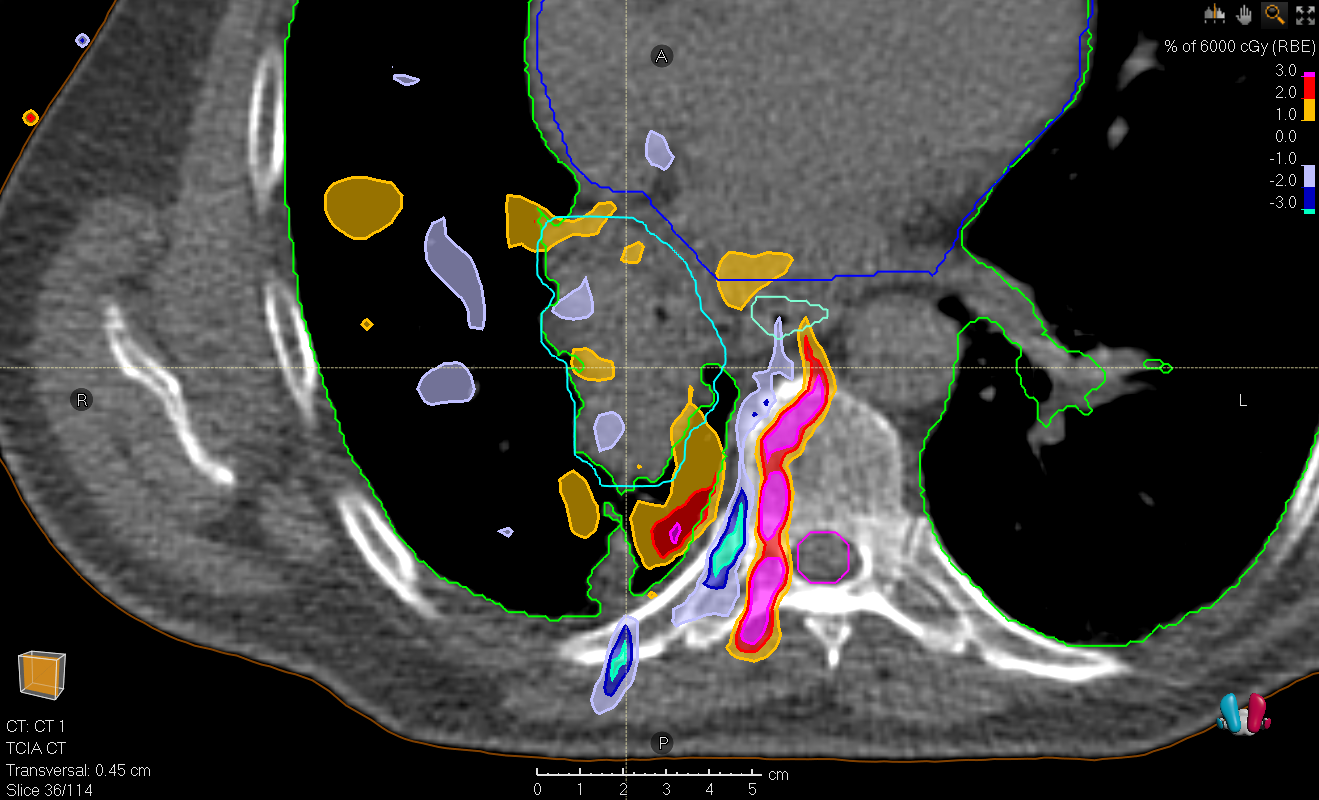}%
\llap{{%  move next graphics to top right corner
\includegraphics[height=2.5cm]{colors.PNG}%
}}
}
\subfloat[$d^{30} - d^{100}$]{%
\includegraphics[trim={3cm 0 4cm 0},clip,width=0.33\textwidth]{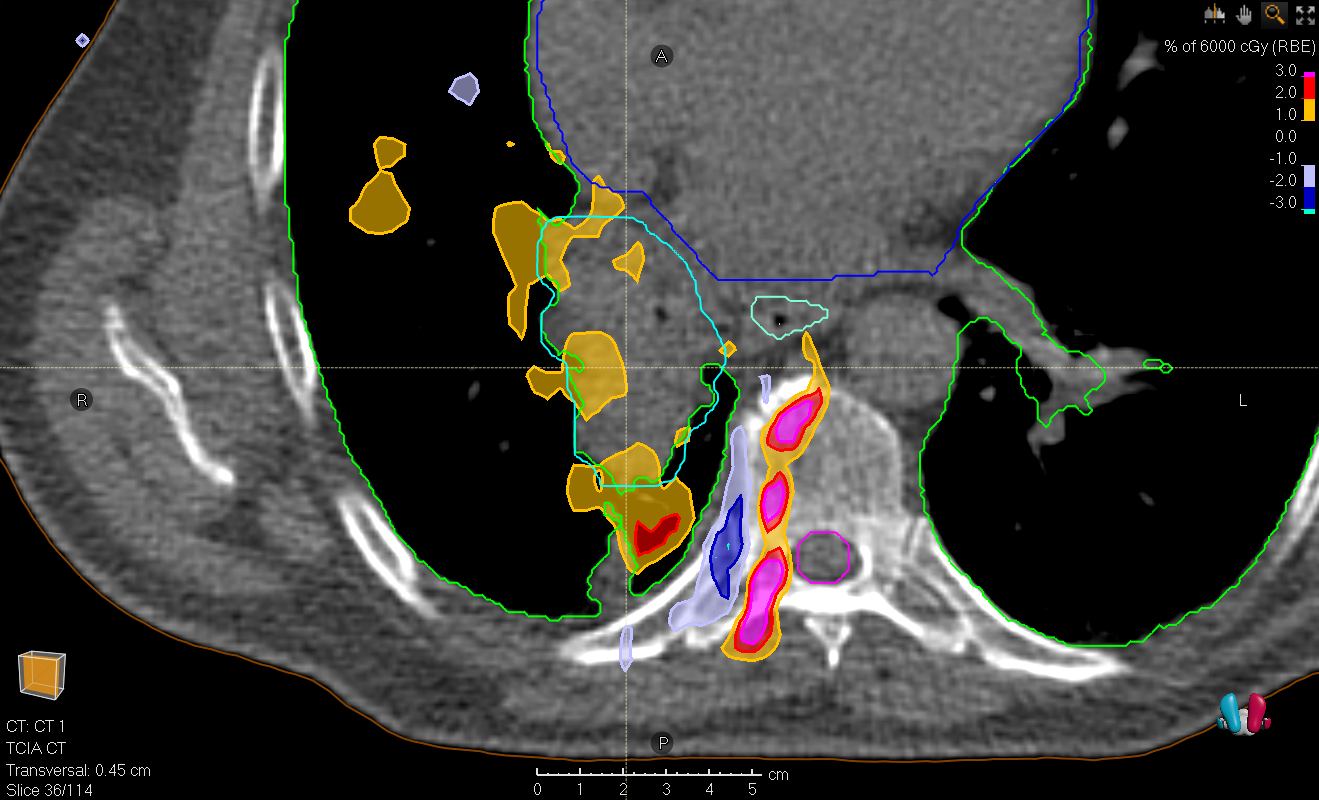}%
\llap{{%  move next graphics to top right corner
\includegraphics[height=2.5cm]{colors.PNG}%
}}
}

\subfloat[$d^{40} - d^{100}$]{%
\includegraphics[trim={3cm 0 4cm 0},clip,width=0.33\textwidth]{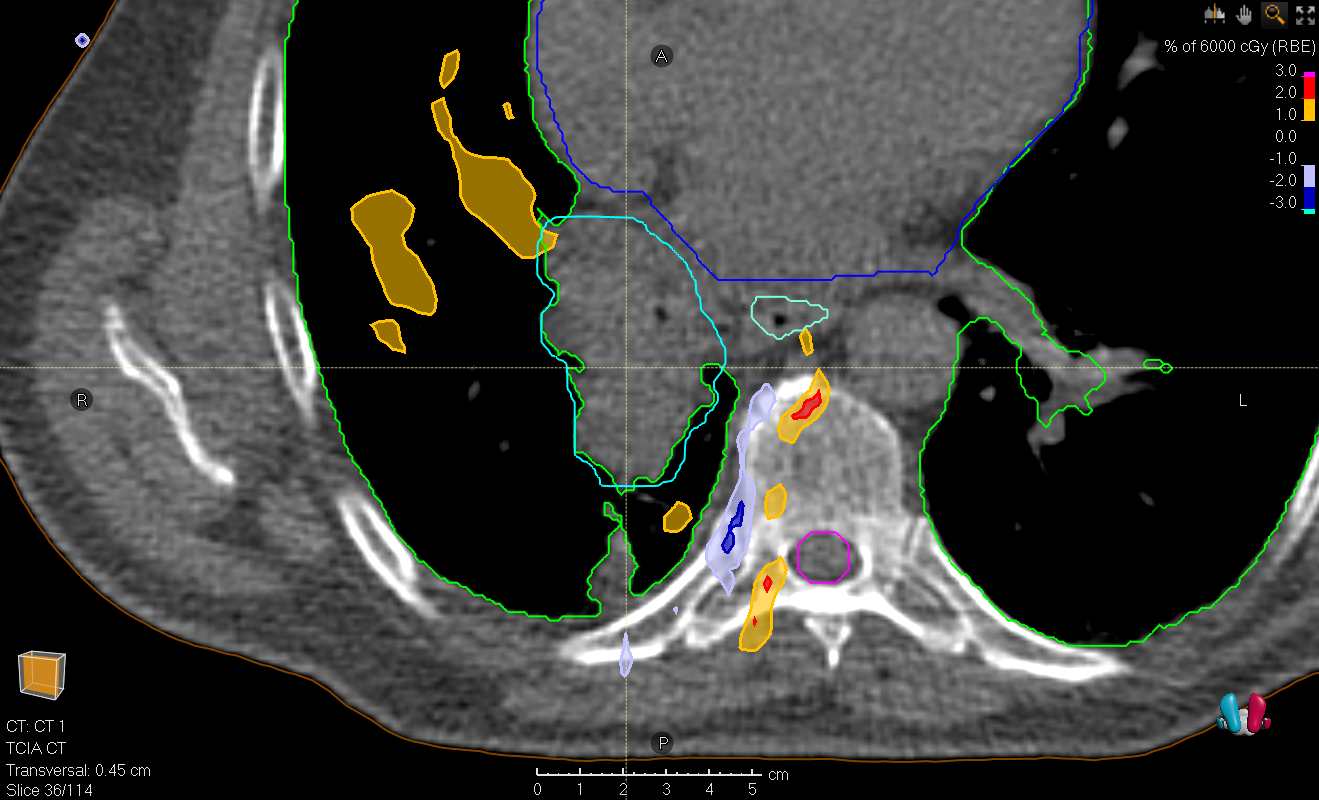}%
\llap{{%  move next graphics to top right corner
\includegraphics[height=2.5cm]{colors.PNG}%
}}
}
\subfloat[$d^{50} - d^{100}$]{%
\includegraphics[trim={3cm 0 4cm 0},clip,width=0.33\textwidth]{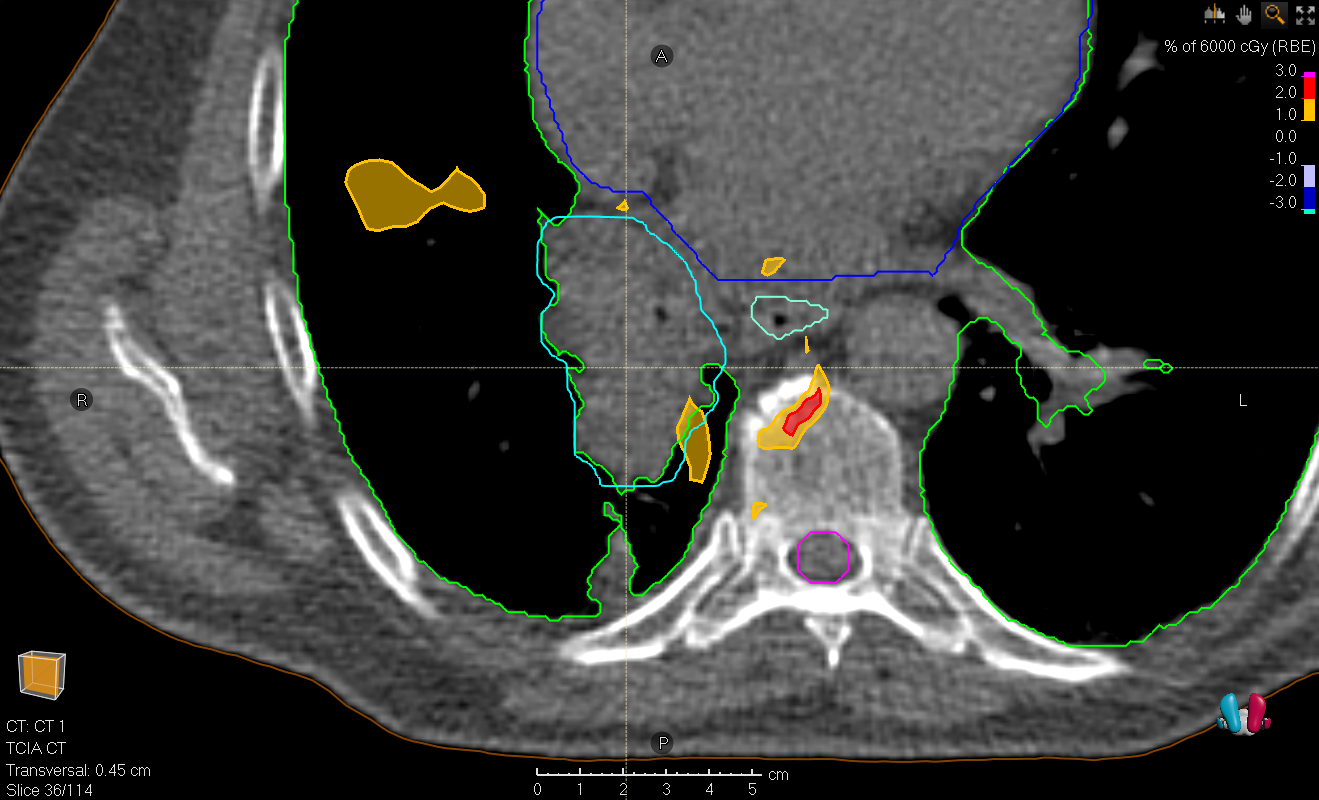}%
\llap{{%  move next graphics to top right corner
\includegraphics[height=2.5cm]{colors.PNG}%
}}
}
\subfloat[$d^{90} - d^{100}$]{%
\includegraphics[trim={3cm 0 4cm 0},clip,width=0.33\textwidth]{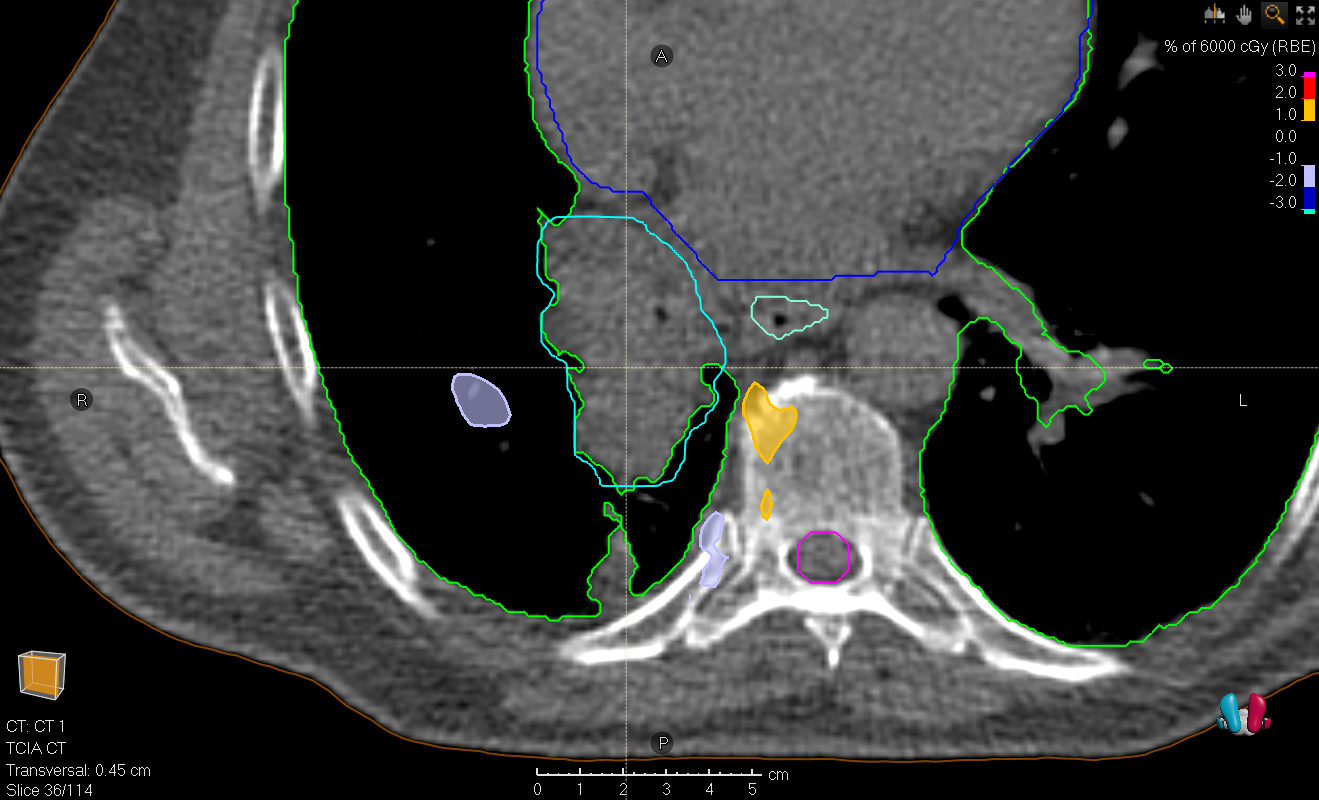}%
\llap{{%  move next graphics to top right corner
\includegraphics[height=2.5cm]{colors.PNG}%
}}
}

\caption{Dose differences (as percentages of the prescribed dose) around the CTV for patient P111. Each subfigure shows the difference from $d^{100}$ for the dose computed at a certain temporal resolution of the 4DCT.}
\label{doses}

\end{figure}

\begin{figure}[]

\subfloat[$l^{10} - l^{100}$]{%
\includegraphics[trim={3cm 0 4cm 0},clip,width=0.33\textwidth]{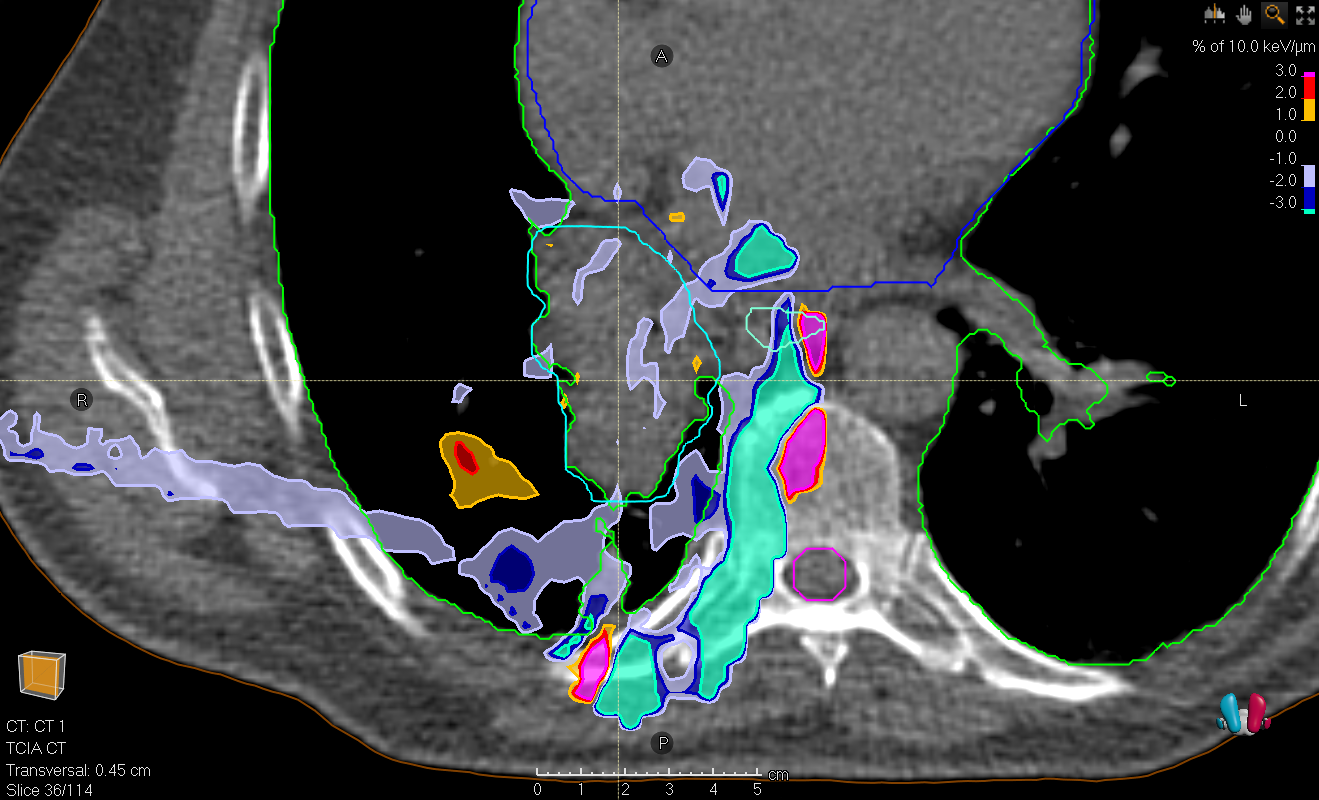}%
\llap{{%  move next graphics to top right corner
\includegraphics[height=2.5cm]{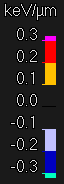}%
}}
}
\subfloat[$l^{20} - l^{100}$]{%
\includegraphics[trim={3cm 0 4cm 0},clip,width=0.33\textwidth]{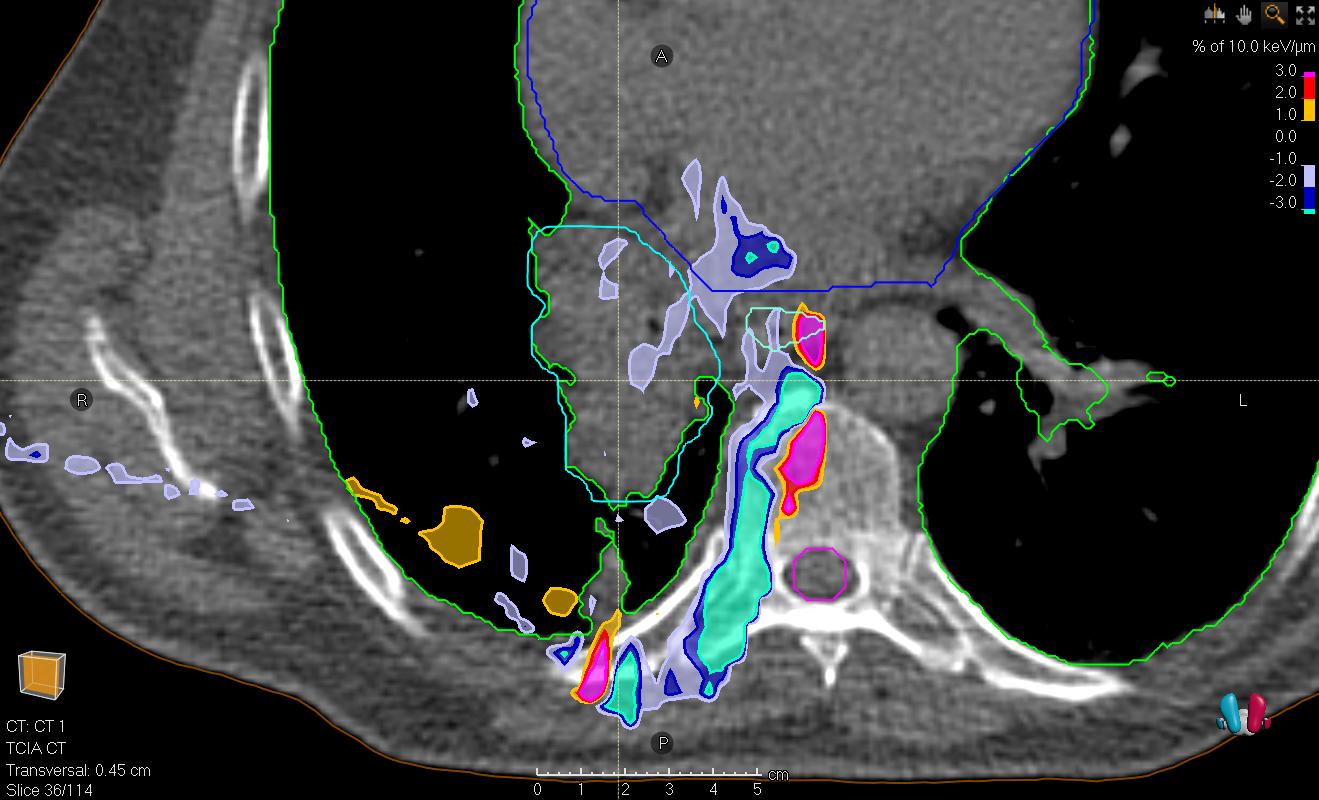}%
\llap{{%  move next graphics to top right corner
\includegraphics[height=2.5cm]{colors2.PNG}%
}}
}
\subfloat[$l^{30} - l^{100}$]{%
\includegraphics[trim={3cm 0 4cm 0},clip,width=0.33\textwidth]{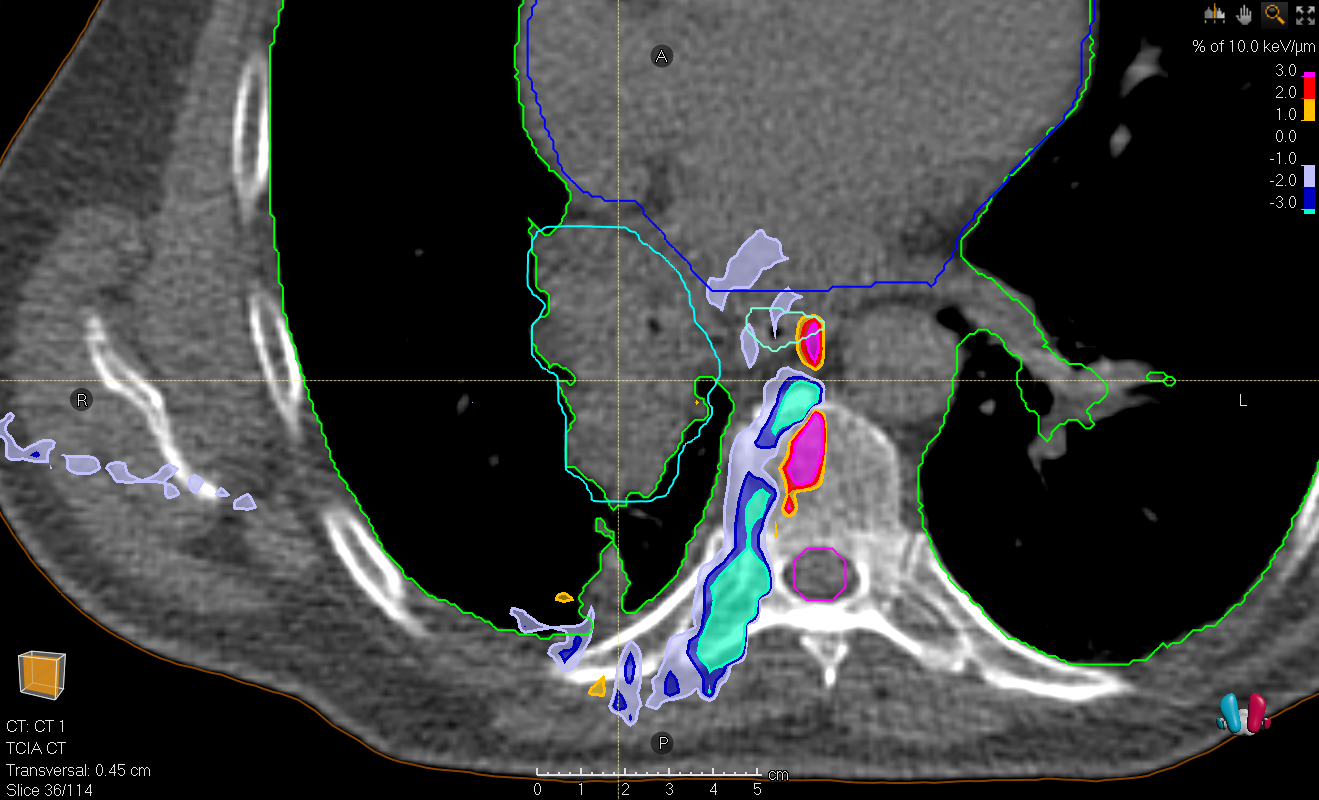}%
\llap{{%  move next graphics to top right corner
\includegraphics[height=2.5cm]{colors2.PNG}%
}}
}

\subfloat[$l^{40} - l^{100}$]{%
\includegraphics[trim={3cm 0 4cm 0},clip,width=0.33\textwidth]{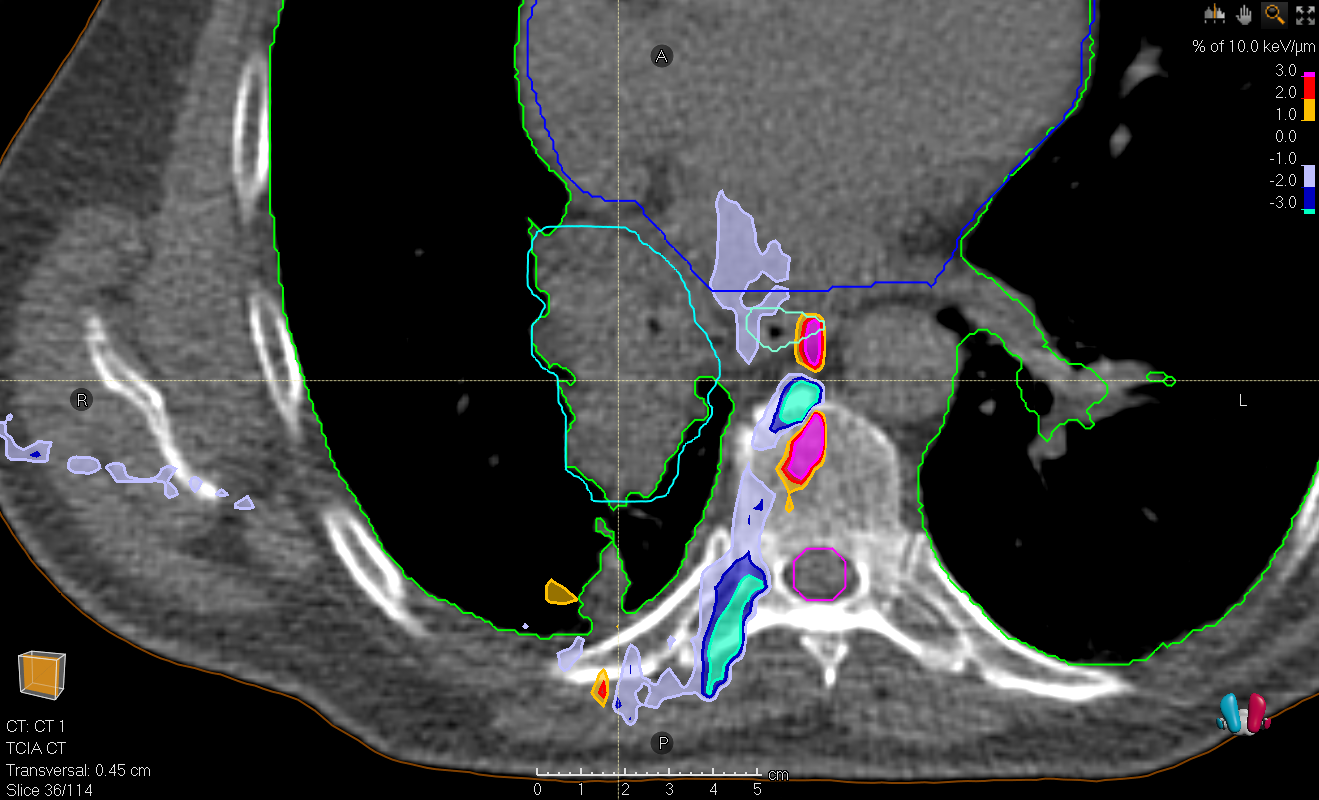}%
\llap{{%  move next graphics to top right corner
\includegraphics[height=2.5cm]{colors2.PNG}%
}}
}
\subfloat[$l^{50} - l^{100}$]{%
\includegraphics[trim={3cm 0 4cm 0},clip,width=0.33\textwidth]{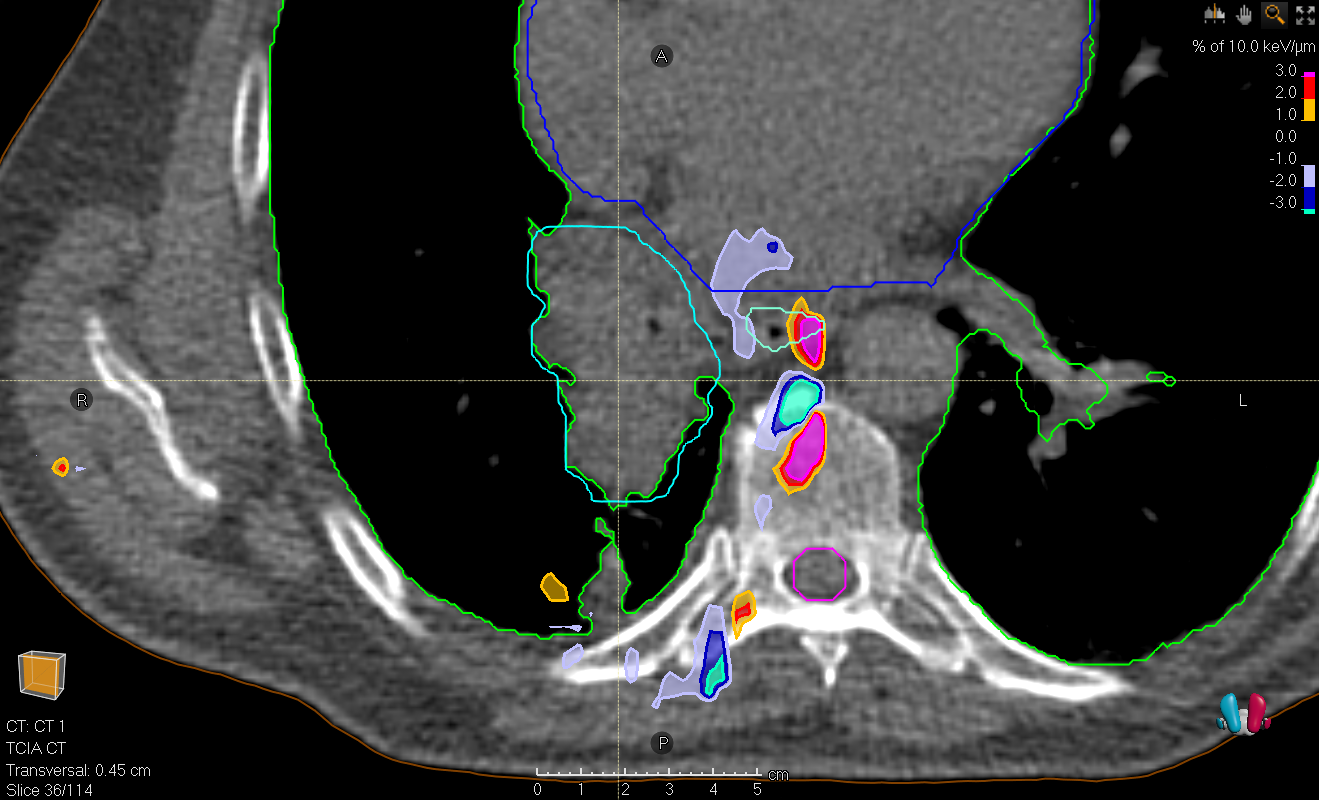}%
\llap{{%  move next graphics to top right corner
\includegraphics[height=2.5cm]{colors2.PNG}%
}}
}
\subfloat[$l^{90} - l^{100}$]{%
\includegraphics[trim={3cm 0 4cm 0},clip,width=0.33\textwidth]{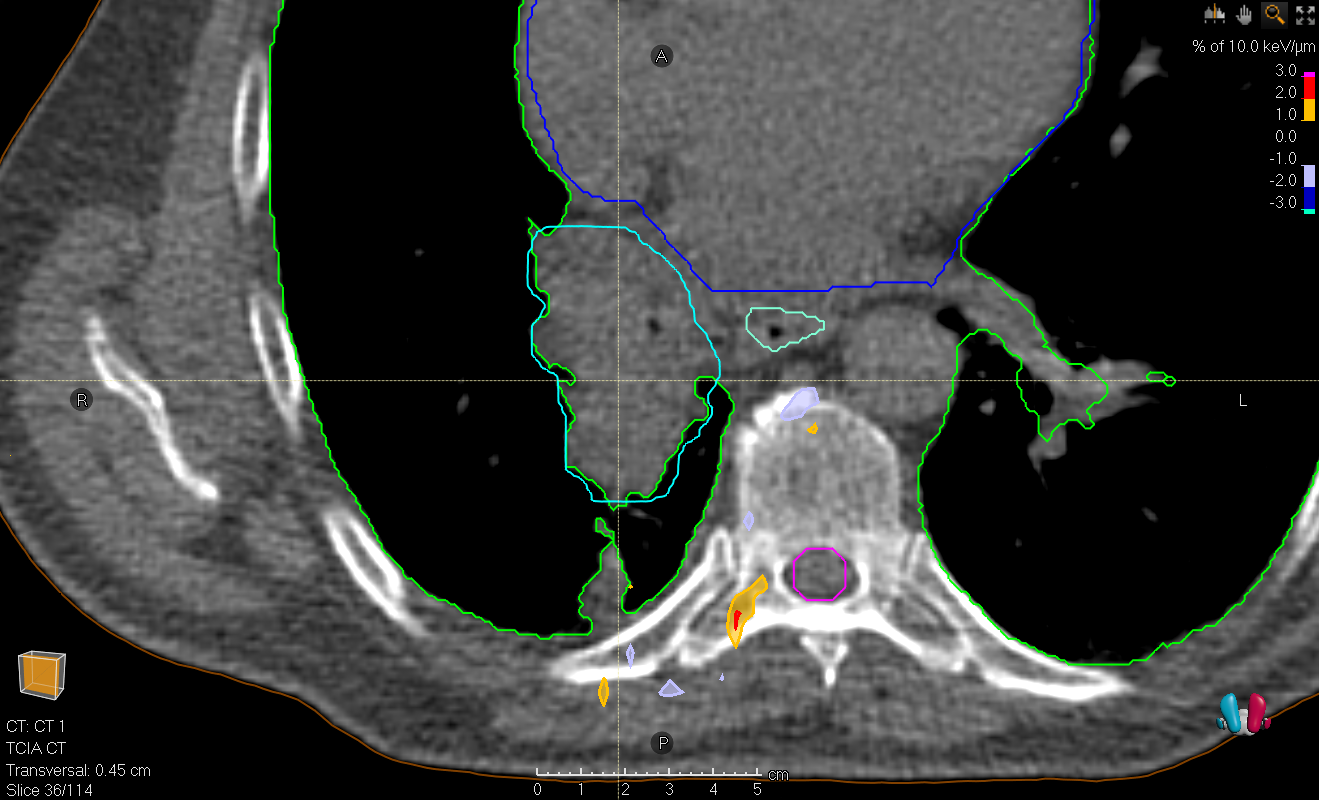}%
\llap{{%  move next graphics to top right corner
\includegraphics[height=2.5cm]{colors2.PNG}%
}}
}

\caption{$\text{LET}_\text{d}$ differences around the CTV for patient P111. Each subfigure shows the difference from $l^{100}$ for the $\text{LET}_\text{d}$ computed at a certain temporal resolution of the 4DCT.}
\label{LETd distributions}

\end{figure}

Finally, for illustrating the convergence to a limit dose distribution, Figure \ref{doses} shows the 4DDC dose distributions for patient P111, with increasing temporal resolution in the 4DCT. Likewise, Figure \ref{LETd distributions} shows the corresponding $\text{LET}_\text{d}$ distributions. 

\subsection{Error severity of 4DDC at low temporal resolution} \label{magnitude}

\begin{figure}[h!]
    \centering
    \includegraphics[trim={4.5cm 0 4.5cm 0},width=\textwidth]{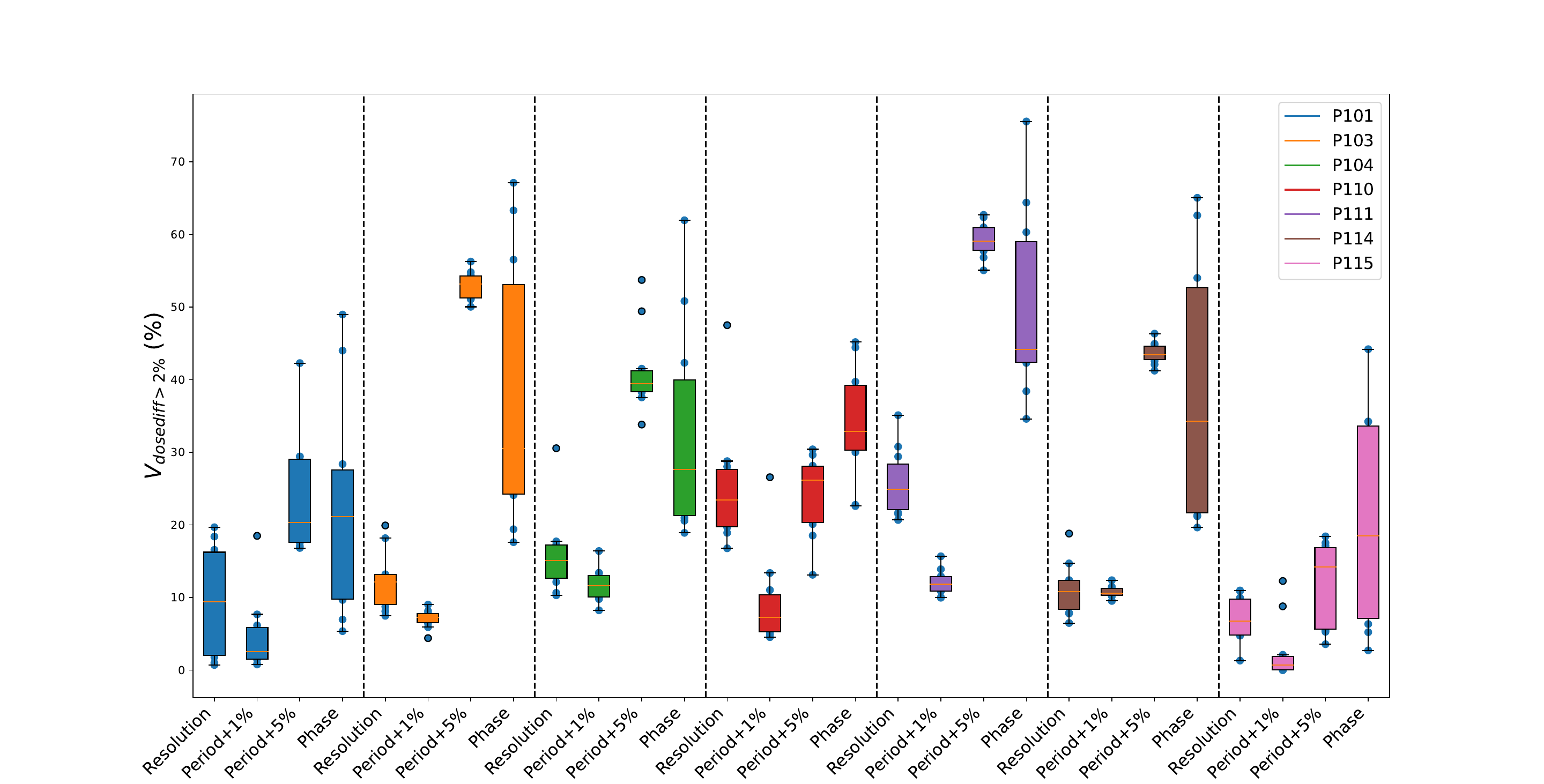}
    \caption{Box plots of the investigated errors for each patient, under varying start phases of the 4DDC. For a specific patient, each plot represents the error source indicated by the x-label. The variation within each box plot was generated by performing the comparison with each of the 10 original phases of the 4DCT used as the start phase of the patient motion in the 4DDC. The dosimetric error is quantified as the $V_{\text{dosediff} > 2 \%}$}
    \label{error_comparison}
\end{figure}

The results in Section \ref{convergence} were based on 4DDCs which all assumed the same start phase of the patient motion. Here, the variation within the results was investigated by comparison of dose differences when different start phases were used. For this purpose, doses computed at the highest and lowest resolutions, $d^{100}$ and $d^{10}$, given the same start phase, were compared against each other, for each of the 10 choices of start phase from the original 4DCT (10 data points). For reference, the errors were juxtaposed with the errors induced by an offset of the start phase by one, as well as errors induced by a change in the patients breathing period by $1 \%$ and $5 \%$. The box plots in Figure \ref{error_comparison} show that the resolution-induced error (measured as $V_{\text{dosediff} > 2 \%}$) was small when compared to those induced by start phase offsets and breathing period increases of $5 \%$. However, the mean resolution-induced error is greater than that corresponding to breathing period increases of $1 \%$.

\section{Discussion}

From all metrics on the right of Figure \ref{convergence_figures}, it can be seen that the differences between consecutive resolution levels decrease as the resolution increases. In particular, the differences measured using the less strict measures are almost negligible for resolutions beyond 20--30 phase images. This is an indication of convergence to a limit dose distribution of the 4DDC, where the approximation error induced by the nearest-neighbour interpolation in the phase sorting approaches zero, and/ or is dominated by other error sources. In addition, the stricter dose metrics exhibit the same convergent behavior for the four patients P103, P104, P111 and P114.

The three remaining patients: P101, P110 and P115, exhibit less apparent convergence results across all comparisons. After further inspection of the corresponding 4DCTs, we hypothesize that this results from certain unexpected irregularities in the image data for these patients. In particular, consecutive phase images exhibit inconsistent positioning of the ribs, which we illustrate in Appendix \ref{appendixA}. In these cases, the deformation fails and the method is not able to create an extended 4DCT which accurately represents the in-between phases. It is plausible that a thoughtful selection of controlling ROIs, or other parameters in the DIR algorithm, on a case-by-case basis, could mitigate this limitation.

The presented results convey two insights related to temporal image resolution when performing 4DDCs for IMPT. The first, deduced from the results in Section \ref{convergence}, is that the interpolation error induced by limited temporal resolution in the 4DCT can be mitigated by generation of the intermediate phases by DIR-based interpolation of neighbouring original phases in the 4DCT. For less strict metrics, measuring $\gamma$-pass rates down to $2\%/2$mm, or voxel-wise dose differences surpassing $2\%$ of the prescription, the error is negligible already at 20--30 images in the extended 4DCT. However, stricter metrics continue to improve with increased temporal resolution up to between 40--70 phase images, for the four patients which converge. For the three remaining patients in this study, the temporal resolution is increased without the stricter metrics clearly converging even at 100 images. These overall results agree well with those of Seo et al. \cite{seo_temporal_2017}, given that the increased complexity introduced by heterogeneous density intuitively increases the demands on temporal image resolution. The second insight, related to the results in Section \ref{magnitude}, is that the error induced by the limited temporal resolution is small when compared to other sources of error, which agrees with the conclusions in Duetschler et al. \cite{duetschler_limitations_2022}.

Convergence to limit $\text{LET}_\text{d}$ distributions in Figure \ref{LETd} was not observed to the same extent as for dose in Figure \ref{convergence_figures}. Although the $\text{LET}_\text{d}$ differences showed a clear decrease with increasing numbers of images, differences were non-negligible even at the highest resolutions. It is plausible that propagation of residual errors from dose and LET computations on individual phase images, as well as deformable registrations, is more severe because of the multiplicative nature of Equation \ref{LETd_formula} for accumulating $\text{LET}_\text{d}$.

We have investigated the impact of limited temporal resolution for phase-sorting-based 4DDC under idealized conditions. The patient motion model assumes known, regular breathing at constant period. As concluded by Duetschler et al. \cite{duetschler_limitations_2022}, other error sources, such as irregular breathing, are severe and introduce additional complexity to the problem. Nevertheless, we believe that our results apply also to irregular patient motion, although confirming this hypothesis requires further research. Another underlying assumption of the method is that the patient motion occurs at a constant rate and direction across all corresponding pairs of points in neighboring phases, along the vectors that align them. In other words, after a fraction $\alpha$, the mass in any point $y_i$ has moved to the point $y_i + \alpha v_i$. This assumption is motivated by the monotonic nature of motion between phases, similarly to as in Zhang et al. \cite{zhang_dosimetric_2019}. Yet another limitation is the inevitable inaccuracy of the deformation itself. Image registrations are inherently uncertain due to limitations of the model. For example, assumptions on the deformation vector field to be smooth and invertible constrain the registration from accurately modelling non-smooth or singular deformations, which may result from changes of mass or structural inconsistencies between the images. In addition, the many degrees of freedom of the registration may lead to uncertainty in regions of low image contrast \cite{brock_use_2017}. Thus, although the interpolation method removes most discontinuity associated with moving between neighbouring phases, some still remains between $R$ and $I_\alpha$ even as $\alpha \to 1$. In addition, the additional deformable registrations needed for dose accumulation on the reference phase in the extended 4DCT may also induce additional errors of this kind. Bearing these limitations in mind, we believe that the results are still valid in terms of the insights presented as to how many phase images are needed for accurate phase-sorting-based 4DDC.

Going forward, the findings suggest that 4DDC should ideally be performed on image data of finer temporal resolution than that in conventional 4DCT. Recently, effort has been made to reconstruct images capturing the patient motion during delivery, using surrogate signals or 4D-MRI \cite{duetschler_motion_2023, annunziata_virtual_2023}. Regardless of the specific technique used, such reconstruction requires selecting a temporal resolution with a favorable trade-off between computation time and 4DDC accuracy. We believe that our results can be used to find that balance.

\subsection{Conclusion}

A method was developed to investigate the impact of the temporal resolution of the 4DCT when performing 4DDC using phase sorting. The results show that although the importance of sufficient temporal resolution varies between patients, 20--30 phase images were required to mitigate the larger resolution-induced errors, while 50 phase images were required when using stricter error metrics on well-behaved data. The patient cohort was limited to seven patients and further application of the method is warranted to draw general conclusions about limitations arising from low temporal resolution in 4DDC.

\subsection{Acknowledgements}

We thank Stina Svensson for valuable input on DIR and Anders Forsgren for valuable discussions and suggestions for the manuscript.

\appendix

\section{Image quality limitations} \label{appendixA}

\begin{figure}
\centering

\subfloat[P101]{%
\includegraphics[clip,width=0.49\textwidth]{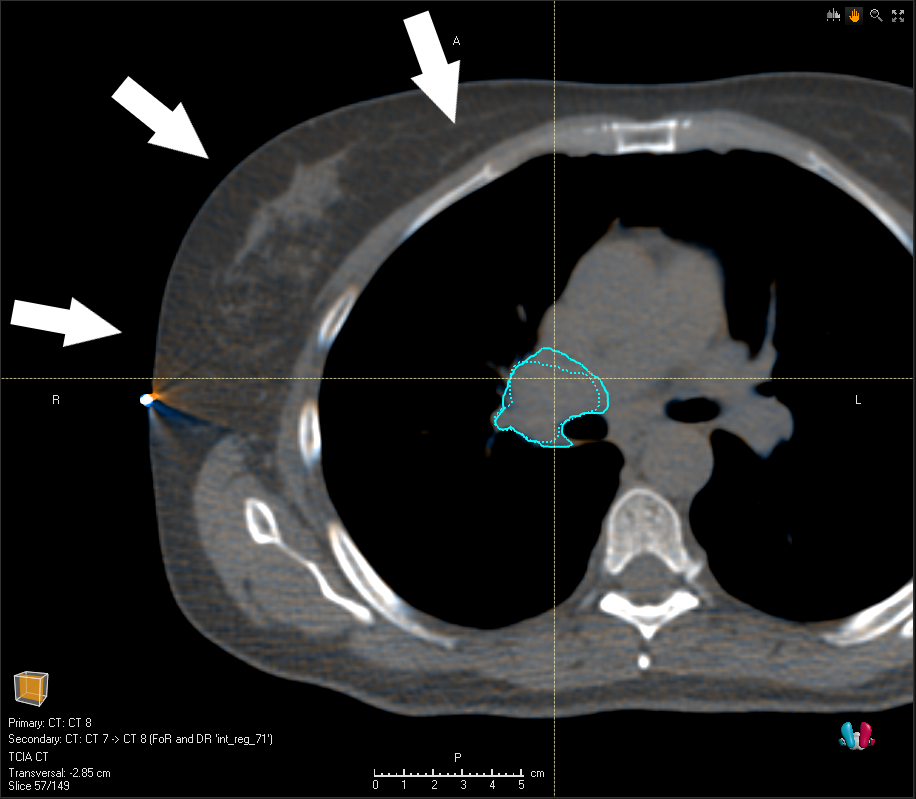}%
}\hfill
\subfloat[P110]{%
\includegraphics[clip,width=0.49\textwidth]{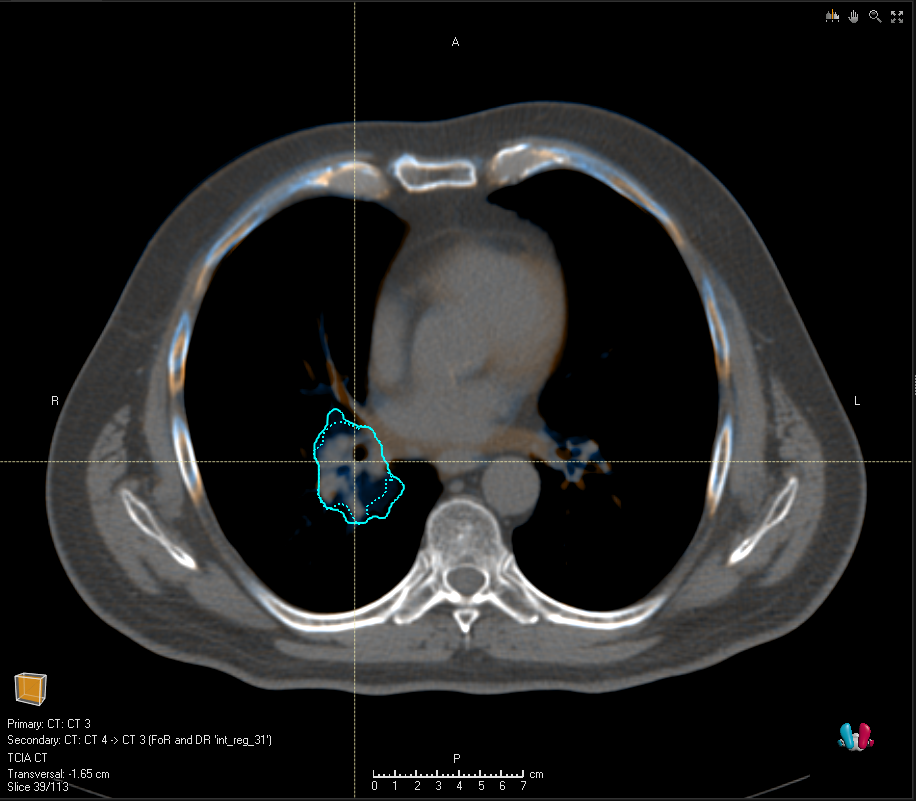}%
}

\subfloat[P115]{%
\includegraphics[clip,width=0.49\textwidth]{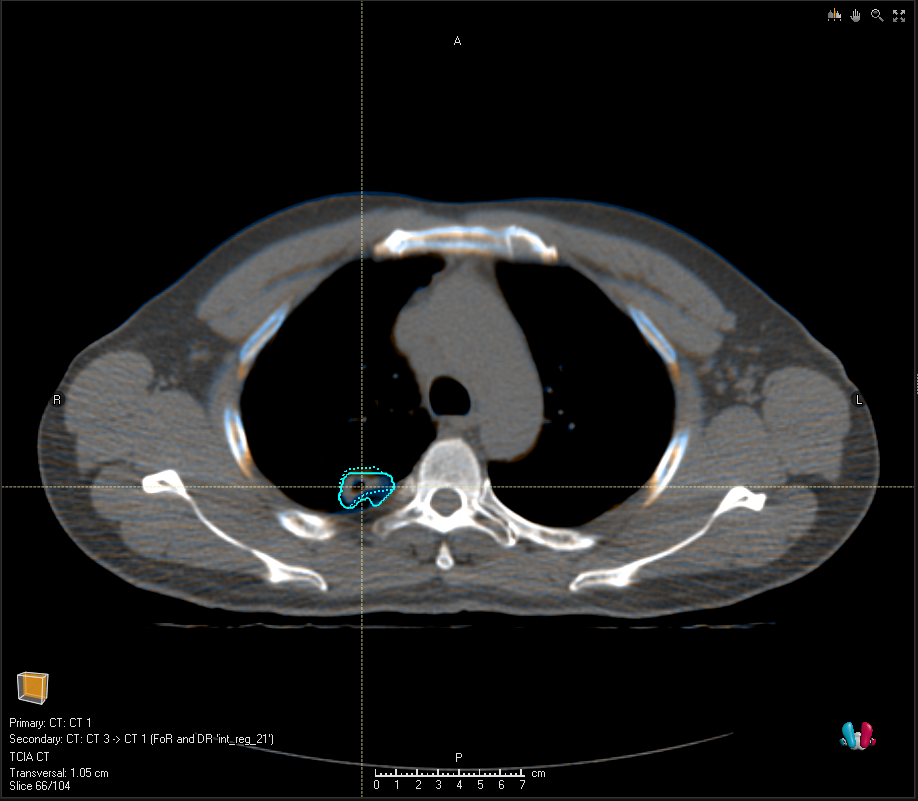}%
}\hfill
\subfloat[P111]{%
\includegraphics[clip,width=0.49\textwidth]{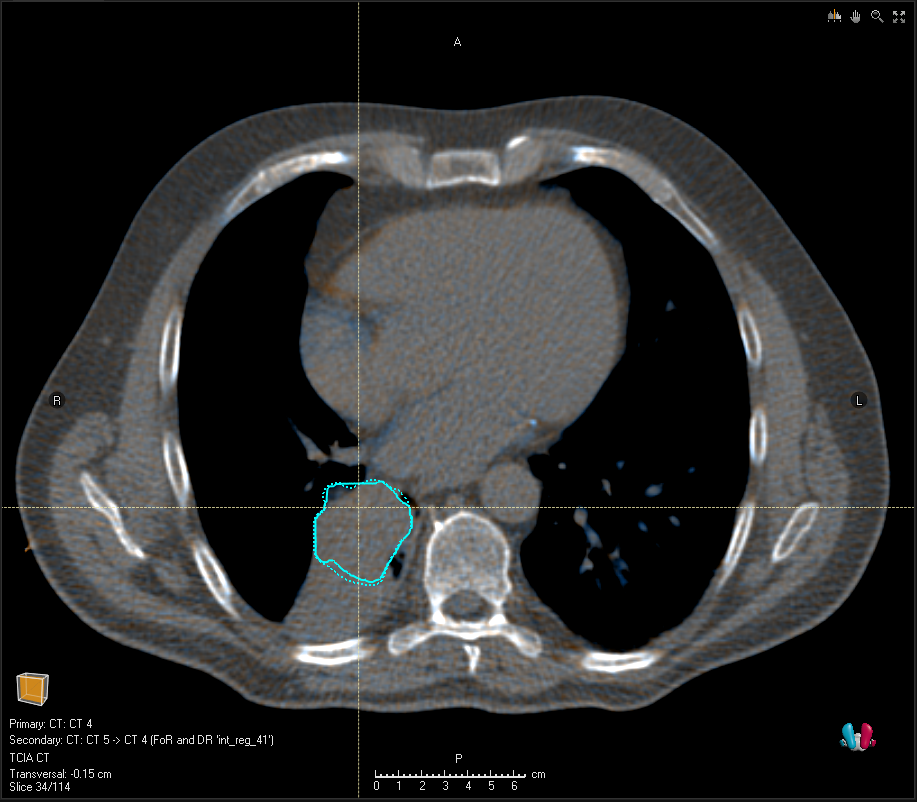}%
}
    
\caption{Illustrations of the failed deformation due to image inconsistencies for patients P101, P110, and P115. Blue and orange regions are indications of deformation failure. Beam angles are indicated by the white arrows.}
\label{bad_ribs}
\end{figure}

As discussed, the resolution induced errors for patients P101, P110, and P115 did not exhibit apparent convergence to zero for stricter error metrics. This observation motivated additional inspection of the 4DCTs for these patients. Upon visual inspection, it was observed that for certain neighbouring pairs of phase images, the deformable registration failed for certain slices. Figure \ref{bad_ribs} showcases example slices from each of the patients in question, as well as an example slice from P111 for reference.

\printbibliography

\end{document}